\documentclass[journal]{IEEEtran}

\def\BibTeX{{\rm B\kern-.05em{\sc i\kern-.025em b}\kern-.08em T\kern-.1667em\lower.7ex\hbox{E}\kern-.125emX}}
\usepackage[numbers]{natbib}
\usepackage{url,hyperref}
\usepackage[switch]{lineno} 
\usepackage{setspace}
\usepackage{graphicx}

\usepackage{indentfirst}
\usepackage{array,multirow}
\usepackage{times} 
\usepackage{fancyhdr}
\usepackage{booktabs}
\usepackage{multirow}
\usepackage{comment}
\usepackage{siunitx}
\usepackage{multicol}
\usepackage{here}
\usepackage{caption}
\usepackage{CJKutf8}
\usepackage[labelformat=simple]{subcaption}

\usepackage{algorithmicx,algorithm,algpseudocode}
\usepackage{amsmath,amsfonts,bm}
\usepackage{xcolor}
\usepackage{fontawesome5}

\usepackage[figuresright]{rotating}

\usepackage[normalem]{ulem}
\useunder{\uline}{\ul}{}

\newcommand{\ctext}[1]{\raise0.2ex\hbox{\textcircled{\scriptsize{#1}}}}

\newcommand{\donebox}{\makebox[0pt][l]{$\square$}\raisebox{.15ex}{\hspace{0.1em}$\checkmark$}}%
\newcommand{\nonbox}{\makebox[0pt][c]{$\square$}}%

\newcommand{\eg}{e.\,g.,\ }
\newcommand{\ie}{i.\,e.,\ }

\usepackage{float}

\begin{document}

\title{An Educational Human Machine Interface Providing Request-to-Intervene Trigger and Reason Explanation for Enhancing the Driver's Comprehension of ADS's System Limitations}

\author{Ryuji~Matsuo, Hailong~Liu$^{*}$,~\IEEEmembership{Senior Member,~IEEE},\\ Toshihiro~Hiraoka,~\IEEEmembership{Member,~IEEE} and Takahiro~Wada,~\IEEEmembership{Member,~IEEE}

\thanks{Ryuji~Matsuo, Hailong Liu and Takahiro Wada are with Graduate School of Science and Technology, Nara Institute of Science and Technology, 8916-5 Takayama-cho, Ikoma, Nara, 630-0192, Japan. }
\thanks{Toshihiro~Hiraoka is with Japan Automobile Research Institute, 1-1-30, Shibadaimon, Minato-ku, Tokyo 105-0012, Japan. }

\thanks{*CONTACT Hailong Liu. \faIcon[regular]{envelope}~:~{\tt\small liu.hailong@is.naist.jp}}
}

\maketitle

\IEEEpubidadjcol

%\linenumbers 

\begin{abstract}
Level 3 automated driving systems (ADS) have attracted significant attention and are being commercialized. 
A level 3 ADS prompts the driver to take control by issuing a request to intervene (RtI) when its operational design domains (ODD) are exceeded.
However, complex traffic situations can cause drivers to perceive multiple potential triggers of RtI simultaneously, causing hesitation or confusion during take-over. 
Therefore, drivers need to clearly understand the ADS's system limitations to ensure safe take-over. 
This study proposes a voice-based educational human machine interface~(HMI) for providing RtI trigger cues and reason to help drivers understand ADS's system limitations. 
The results of a between-group experiment using a driving simulator showed that incorporating effective trigger cues and reason into the RtI was related to improved driver comprehension of the ADS's system limitations. 
Moreover, most participants, instructed via the proposed method, could proactively take over control of the ADS in cases where RtI fails; meanwhile, their number of collisions was lower compared with the other RtI HMI conditions.
Therefore, using the proposed method to continually enhance the driver's understanding of the system limitations of ADS through the proposed method is associated with safer and more effective real-time interactions with ADS.

\end{abstract}

\section{Introduction}
\label{sec:Introduction}

\IEEEPARstart{A}{}utomated Driving System~(ADS) has been a growing interest in recent years~\cite{9046805,li2021autonomous}.
The Society of Automotive Engineers (SAE) classifies ADS into levels 0 to 5~\cite{sae2018taxonomy}.
In which, the level 3 ADS is conditional driving automation where the driver is not required to continuously monitor the driving environment~\cite{sae2018taxonomy}. 
%During the use of level 3 ADS, the driver can engage in non-driving related activities (NDRT).
However, the driver must take over control of the vehicle in response to a request-to-intervene~(RtI) from the Level 3 ADS as it approaches the limits of its operational design domain~(ODD) or system limitations.
Once the RtI is issued, the ADS will disengage if the ODD or system limitations are surpassed. Consequently, the prompt reaction of the driver to the RtI is essential to guarantee safety.

\subsection{Take-over Issues In Using Level~3 ADS}
\label{sec:Issues in take-over}

In general, when the ADS issues an RtI, the driver may take approximately 10 seconds to respond, followed by an additional 25-30 seconds to regain stable control of the vehicle~\cite{merat2014transition, strand2014semi}. 
This delay can be attributed to the driver needing to regain situational awareness, assess risk, and decide how to take over vehicle control~\cite{saito2018control,kondo2019use, okada2020transferring}.
Additionally, after taking over the control, the driver also needs time to re-adapt to vehicle dynamics and road conditions in order to smoothly control the vehicle.

\begin{figure}[tb]
  \centering
  \includegraphics[width=1\linewidth]{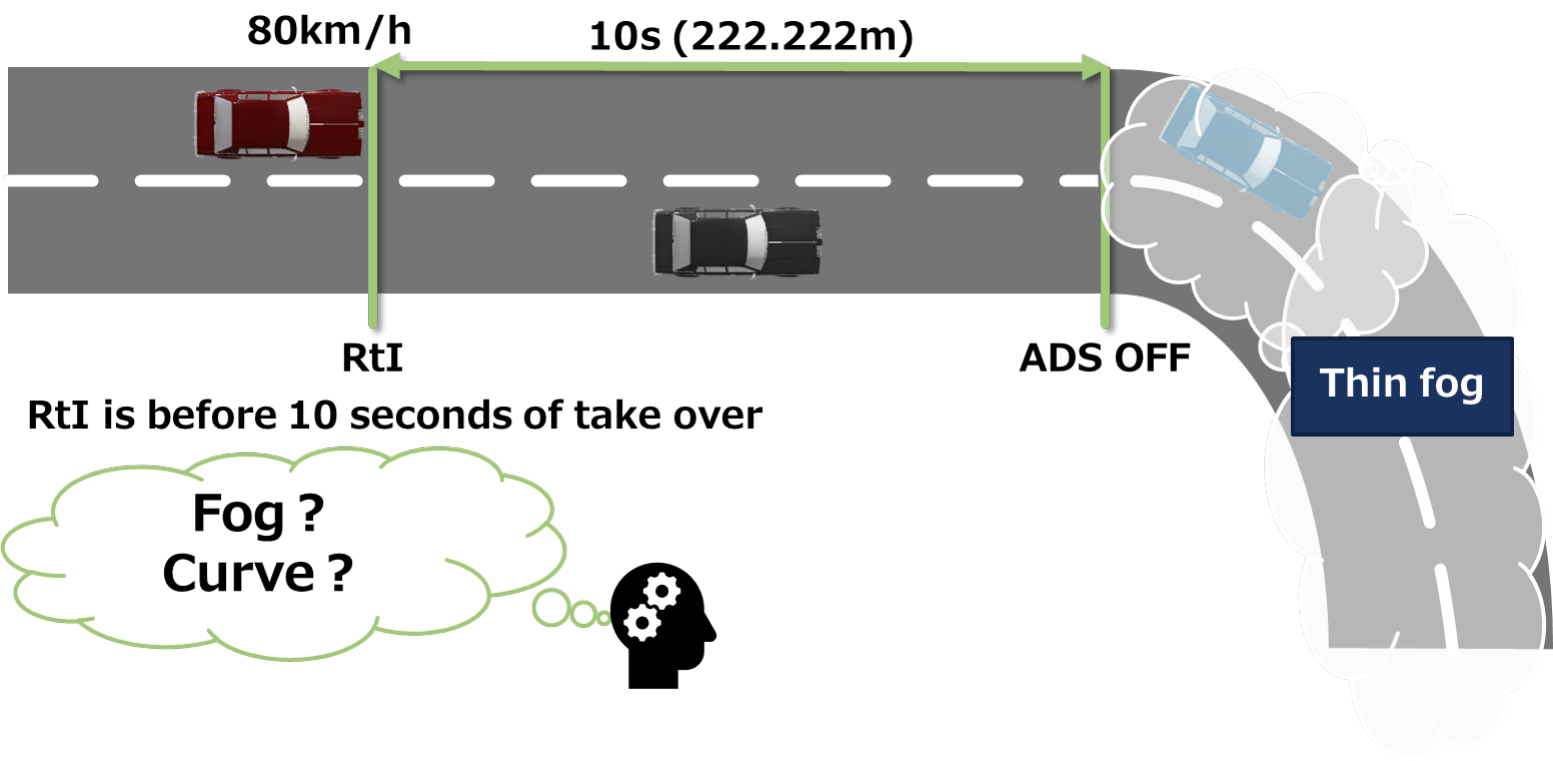}
  %\vspace{-4mm}
  \caption{The ADS issues an RtI due to the sharp curve ahead rather than the thin fog. In such a complex traffic scenario, the driver may encounter challenges in identifying the cause of the RtI trigger.}
  \label{fig:scenario_ex}
\end{figure}

Furthermore, in complex scenarios, drivers may encounter situations where more than one potential trigger seems present.
For example, Fig.~\ref{fig:scenario_ex} illustrates a case in which an AV issues an RtI due to a sharp curve ahead, while the environment is also affected by thin fog.
Although the RtI is triggered only by the sharp curve, the coexistence of these factors may cause drivers to perceive multiple simultaneous triggers, leading to confusion.
As a result, drivers must first determine the actual cause of the ADS's RtI before taking action, and any misinterpretation may result in hesitation or inappropriate responses.
Such uncertainty can make drivers hesitate over whether to steer or brake first, leading to delayed or incorrect intervention and thereby increasing the risk of accidents~\cite{kondo2019shared}.
Responding to RtIs under such conditions thus poses significant challenges for the driver's situational awareness and risk assessment.
In addition, over-trusting the ADS presents another potential risk since drivers who misunderstand its limitations and misinterpret the triggers for RtI may also fail to respond appropriately, which can result in accidents~\cite{liu2019overtrust}.

To address the above issues, we consider that the driver needs to 1) understand potential challenges the ADS may encounter, 2) recognize the trigger of the RtI protocol based on the surrounding conditions, and 3) take appropriate actions during the take-over.
This requires the driver to cognitively comprehend the limitations and capabilities of the ADS.
In essence, to safely take control of the vehicle during an RtI, the driver needs to establish a correct mental model of the ADS. 
The mental model represents the driver's knowledge of the ADS mechanism and its limitations~\cite{liu2019driving,zhou2021does}, which can be developed through repeated usage~\cite{staggers1993mental} and appropriate education~\cite{zhou2021does,liu2021importance,liu2025}.

\subsection{Related Works}
\label{sec:RelatedWorks}

To help drivers effectively respond to the RtI of the level 3 ADS, two main approaches can be considered: 1) human machine interface~(HMI) for take-over assistance; 2) education to calibrate the mental model of the ADS.
We consider that the take-over assistance HMIs can enhance the driver's situational awareness when an RtI is issued.
Additionally, the education can help drivers understand the system limitations and rationales of ADS, thereby assisting in predicting the operating status of the ADS.
Relevant studies on these two approaches are investigated as follows.

\subsubsection{Human machine interfaces for take-over assistance}
\label{sec:HMI for take-over assistance}

Many studies have proposed various HMIs based on auditory, visual, and haptic cues to assist drivers in taking over control.
Haptic cues have been used to help drivers take over the control safely.
\citet{schwalk2015driver} evaluated the effectiveness of using a vibrotactile seat matrix with different vibration patterns to cue the RtI.
Besides, \citet{borojeni2017comparing} proposed a shape-changing steering wheel that the grip surfaces were moving under drivers' palms to hint the recommended steering direction after the RtI was issued.
The results indicated that haptic cues on the steering wheel during RtI serve to reassure drivers about their decisions rather than assist them in decision-making. 
However, this method falls short in conveying the rationale behind the RtI, as it lacks the capability to provide detailed information.
Moreover, \citet{morales2022automated} proposed a haptic guidance during the take-over to help the driver avoid surrounding obstacles by applying a reaction force to the steering wheel.
While this force guidance helps in reducing human errors during take-over, it fails to communicate to the driver the current situations and the rationale for the RtI being issued.

Various auditory cues have been extensively used for RtI.
For example, \citet{Politis2015} and \citet{forster2017driver} compared speech-based warning with beep sound warning.
They found that combining beep sound warnings with speech-based warnings resulted in better subjective feelings than using beep sound warnings alone, but there was no significant difference in response time between the two audible warning conditions. 
However, we think that while the proposed sound-based warning RtI HMI can enhance driver alertness, it remains challenging to improve drivers' awareness of the driving environment and their understanding of the occurring specific events.

Visual information cues are often combined with auditory cues for RtI.
\citet{el2023influence} proposed a color themed HMI which uses ambient light and auditory cue to send RtI to the driver. 
Its results indicate that compared to the auditory HMI, the color-themed HMI was related to better driver understanding of driving tasks and shorter RtI reaction times.
Similarly, \citet{gonccalves2023users} also proposed an ambient light-based HMI to issue the RtI.
They found that ambient light not only provided RtI information but also made drivers more easier to focus their attention on the road.
\citet{ou2021effects} also used a combination of visual and auditory HMI, which not only alerted the RtI but also provided a countdown to the ADS turn OFF and indicated the direction of the RtI trigger.
We consider that while the above HMIs effectively convey RtI information, they do not provide drivers with information about the RtI triggers, which is necessary to improve situational awareness during take-over.

To improve drivers' situation awareness when the RtI issued, 
\citet{Borojeni2016} used an LED bar to not only signal RtI but also to convey contextual information, such as the presence of roadblocks. 
They indicated that providing contextual cues during RtI prompts drivers to check their surroundings, leading to a decrease in reaction time and an increase in the time-to-collision (TTC).
Moreover, \citet{wright2018effective} proposed an auditory cues regarding upcoming hazards, such as ``crosswalk ahead'' for environmental cue and ``crosswalk ahead, scan for pedestrians'' for threat cue.
Their findings suggested that environmental cues could increase the likelihood of drivers avoiding crashes compared to the threat cue and the warning cue (\ie ``take over control'').
We think that the RtI HMI proposed in study \citep{wright2018effective} provides the trigger of RtI to drivers, which is critically important for enhancing drivers' awareness of the surrounding environment.
However, the RtI triggers such as crosswalk and roadblocks, are binary variables, \eg there is a crosswalk or not, which makes it easy for drivers to understand this system limitation of ADS.
For more complex triggers involving continuous variables like visibility changes due to fog or rain, providing RtI triggers such as ``fog'' or ``rain'' to drivers may not fully promote understanding of the essential reasons of RtI issuance and the specific visibility threshold of system limitations.
Furthermore, if drivers do not clearly understand the system limitations, they may not be able to proactively take over control when the ADS's RtI system fails to issue alerts.

\subsubsection{Education for calibrating the mental model of the ADS}
\label{sec:Issues in pre-education}

To enable drivers to safely take over driving tasks after an RtI is issued, they need to understand the limitations of the ADS and establish an accurate mental model of the system~\cite{liu2019overtrust}.
The mental model is an internal representation of a target system which contains meaningful declarative and procedural knowledge derived from long-term experiences and studies~\cite{endsley1995toward,jones2011mental}. 
Further, education can be used for drivers to develop a mental model of ADS quickly~\cite{zhou2021does, boelhouwer2019should, EBNALI2019184, liu2021importance,liu2025}.

\citet{boelhouwer2019should} focused on whether the current method of providing information on the ADS to drivers, primarily through user manuals, can align the driver's mental model with the capabilities of the ADS.
As a result, the user manuals did not appear to support the participants in correctly deciding whether to take over or rely on the ADS.
\citet{zhou2021does} focused on using pre-instruction to provide drivers with prior knowledge of the system limitations of level 3 ADS.
Their findings revealed that prior knowledge improved the success rate of take-over and reduced reaction time during the first experience with the RtI.
According to \citet{EBNALI2019184}, simulator training and video training had a more positive effect on take-over time and accuracy.
In addition, the simulator training participants outperformed in deciding whether take-over was necessary or not. 
However, both studies highlight an issue that the learning effect of knowledge is diminished because drivers easily forget it over time after a few RtI experiences.

Providing continuous education on ADS for drivers is an important and challenging task because the mental model of ADS requires drivers to spontaneously develop it through repeated recognition and interpretation~\cite{staggers1993mental}. 
\citet{korber2018have} proposed implementing a post hoc explanation 14 s after the presentation of the RtI to improve the trust and acceptance of the ADS by drivers.
Their explanation content focuses on the cause and effect of RtI, but does not explain the reasons associated with the system limitations of the ADS.
After multiple repetitions of the take-over events, they found that the post hoc explanation did not affect the drivers' trust and acceptance of the ADS, but it increased their perceived understanding of the system.
Regrettably, they did not verify whether the drivers' understanding of the system was correct.

Based on the conclusions of the above related research, our pre-study ~\cite{Matsuo2024ICHMS} proposes an HMI that provides a RtI trigger and a reason explanation.
The pre-study conducted a small-scale participant experiment to compare the proposed HMI with conventional warning-type RTI, and verified the effectiveness of the proposed HMI.
Nevertheless, it remains unclear whether this effectiveness is due to the RtI triggers or the RtI reason explanations.

%Moreover, it can be challenging for drivers to identify the correct trigger and respond appropriately based on their prior knowledge when there are multiple possible triggers of the RtI from the driver's point of view. 

%In such a scene, it can be challenging for drivers to identify the correct trigger and respond appropriately based on their prior knowledge when there are multiple possible triggers of the RtI from the driver's point of view. 
%On the other hand, providing cues to the driver can enhance situational awareness.

\subsection{Purpose}
\label{sec:RQ}

The purpose of this study is to propose a human-machine interface (HMI) designed to continually educate and assist drivers in ensuring that they consistently establish an accurate mental model of the ADS, understand the system limitations, and make correct inferences about triggers and reasons for RtI, particularly in scenarios where multiple triggers may appear ambiguous from the driver's perspective.

Based on our pre-study~\cite{Matsuo2024ICHMS}, this study conducted a larger-scale participant experiment and included an additional comparison condition referring~\citep{wright2018effective}, \ie RtI HMI with trigger cue, to further analyze the effectiveness of RtI triggers and RtI reason explanations to help drivers understand the limitations of the system in this paper.

% \subsection{Structure of this thesis}
% \label{sec:Structure}

% This study first surveys relevant research on RtI and describes the RtI problem in Section~\ref{sec:Introduction}. Section ~\ref{sec:Proposed_Method} explains details of proposed methods for RtI. 
% Next, Section~\ref{sec:Experiment} describes the details of the implementation of proposed methods and the experimental setup and evaluation items used to test its effectiveness. 
% Then, Section~\ref{sec:Results} presents the results of this study.
% Section~\ref{sec:Discussion} discusses from the results. 
% Finally, Section~\ref{sec:Conclusion} summarizes this study and discusses future issues and prospects.

% For this purpose, we propose a voice-based instruction HMI that consists of two components: 1) RtI trigger cues to assist drivers in rapidly enhancing their situational awareness, specifically to help drivers correctly identify the correct trigger for the RtI of an ADS among multiple potential triggers; 2) post-event reason explanations to help drivers learn about the system limitations of the ADS.
% For example, in the scenario shown in Fig.~\ref{fig:scenario_ex}, the ADS provides an RtI trigger cue via voice, \eg, ``Thick fog, take over! Thick fog, take over!'' 
% In addition, after the driver turns ON the ADS after the take-over event, the ADS provides an RtI reason via voice, \eg, ``The reason for RtI is that it is difficult to recognize lanes and vehicles ahead in thick fog when it is impossible to see more than 40~m ahead.''  

\section{Proposed Method}
\label{sec:Proposed_Method}

\begin{figure}[t]
  \centering
  \includegraphics[width=\linewidth]{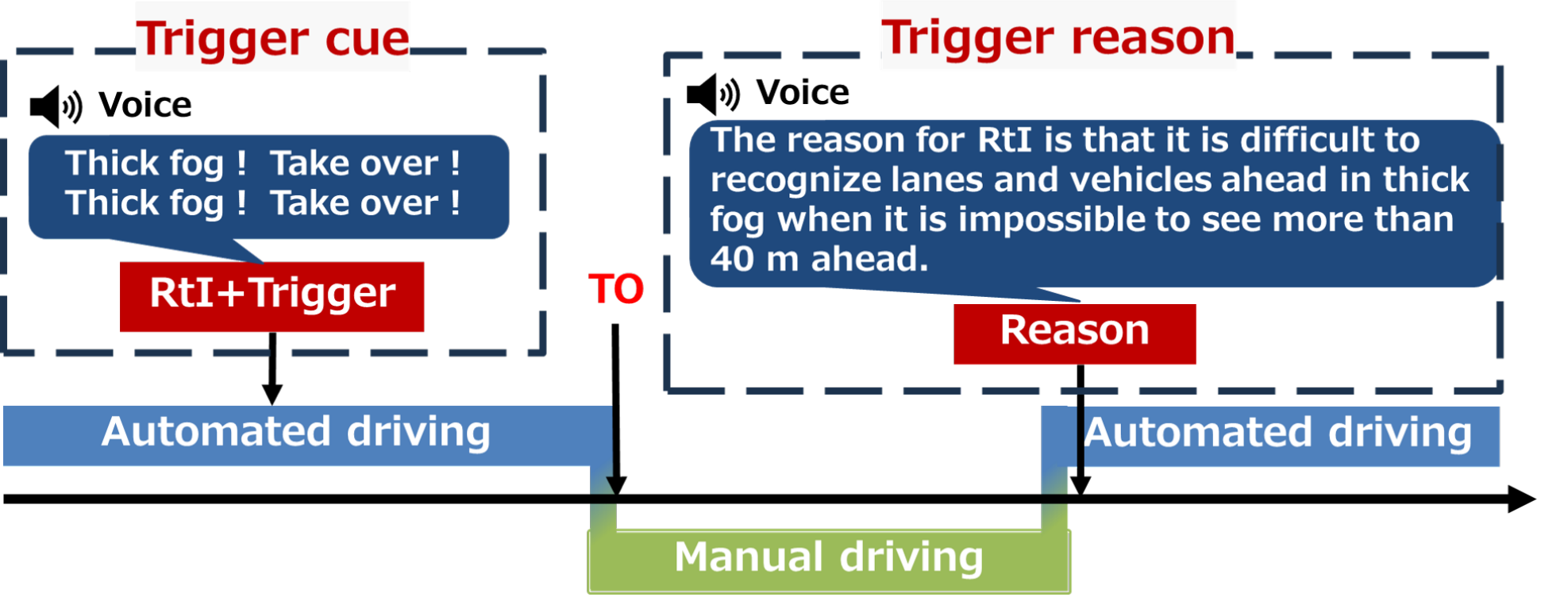}
  \caption{Proposed HMI that provides the driver information about the trigger cues and reasons for the RtI.}
  \label{fig:trigger_reasons}
\end{figure}

This study proposes a voice-based educational HMI for providing trigger cues of RtI and its reasons to assist the driver during take-over as shown in Fig.~\ref{fig:trigger_reasons}.
Voice-based method was primarily used because~\citet{petermeijer2017driver} demonstrated that auditory and tactile take-over requests yielded faster reactions than visual take-over requests.

\subsection{RtI Trigger Cue}

In this study, as shown in Fig.~\ref{fig:scenario_ex}, the RtI is designed to be triggered 10 seconds before the ADS exceeds the system limitations, as referenced in~\cite{merat2014transition}.
Therefore, since the ADS will be deactivated once it exceeds the system limitations, the driver needs to promptly and safely take over vehicle control.

Drivers who have received an RtI need to regain situational awareness as soon as possible to safely resume vehicle control.
Therefore, this study proposes a voice-based HMI that issues trigger cues to assist drivers who have received an RtI to regain situational awareness.
As shown in Fig.~\ref{fig:trigger_reasons}, this method provides comprehensive information concerning trigger cues to enhance situational awareness when the ADS initiates a RtI.
The audio of RtI and trigger cue are designed to be no longer than 4 seconds.
For example, if the visibility range is less than 40~m due to thick fog, the ADS will issue the RtI with the voice cue: ``Thick fog, take over! Thick fog, take over!''.
Drivers who receive trigger cues are believed to be able to reduce their risk of collision, similar to the results of \cite{wright2018effective}. 
This study focuses on drivers' understanding of the system limitations.

\begin{figure*}[h!tb]
\centering
\begin{subfigure}{0.45\linewidth}
  \includegraphics[width=1\linewidth,clip,trim=80 0 0 50]{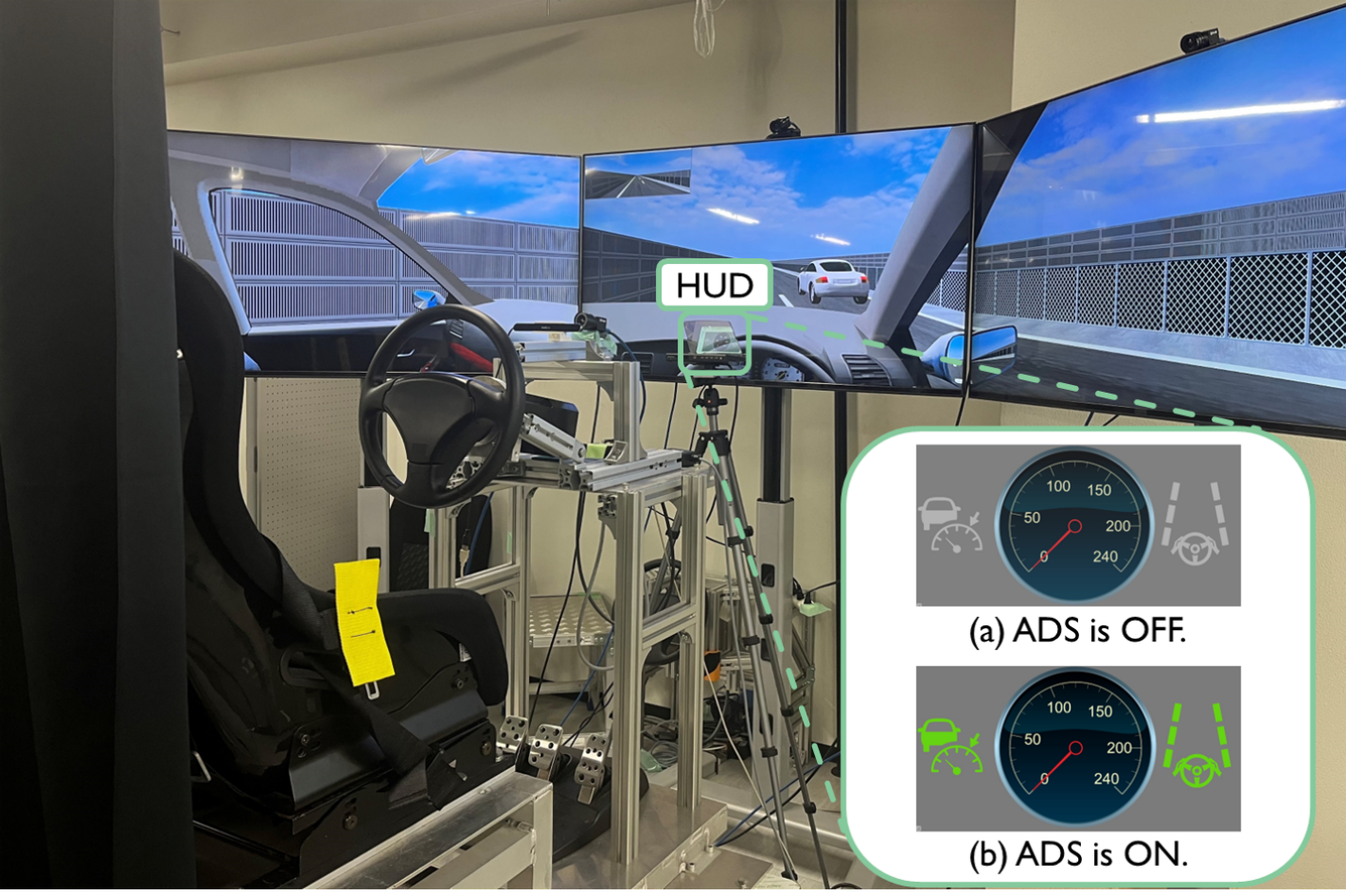}
\end{subfigure}
\begin{subfigure}{0.359\linewidth}
  \includegraphics[width=1\linewidth]{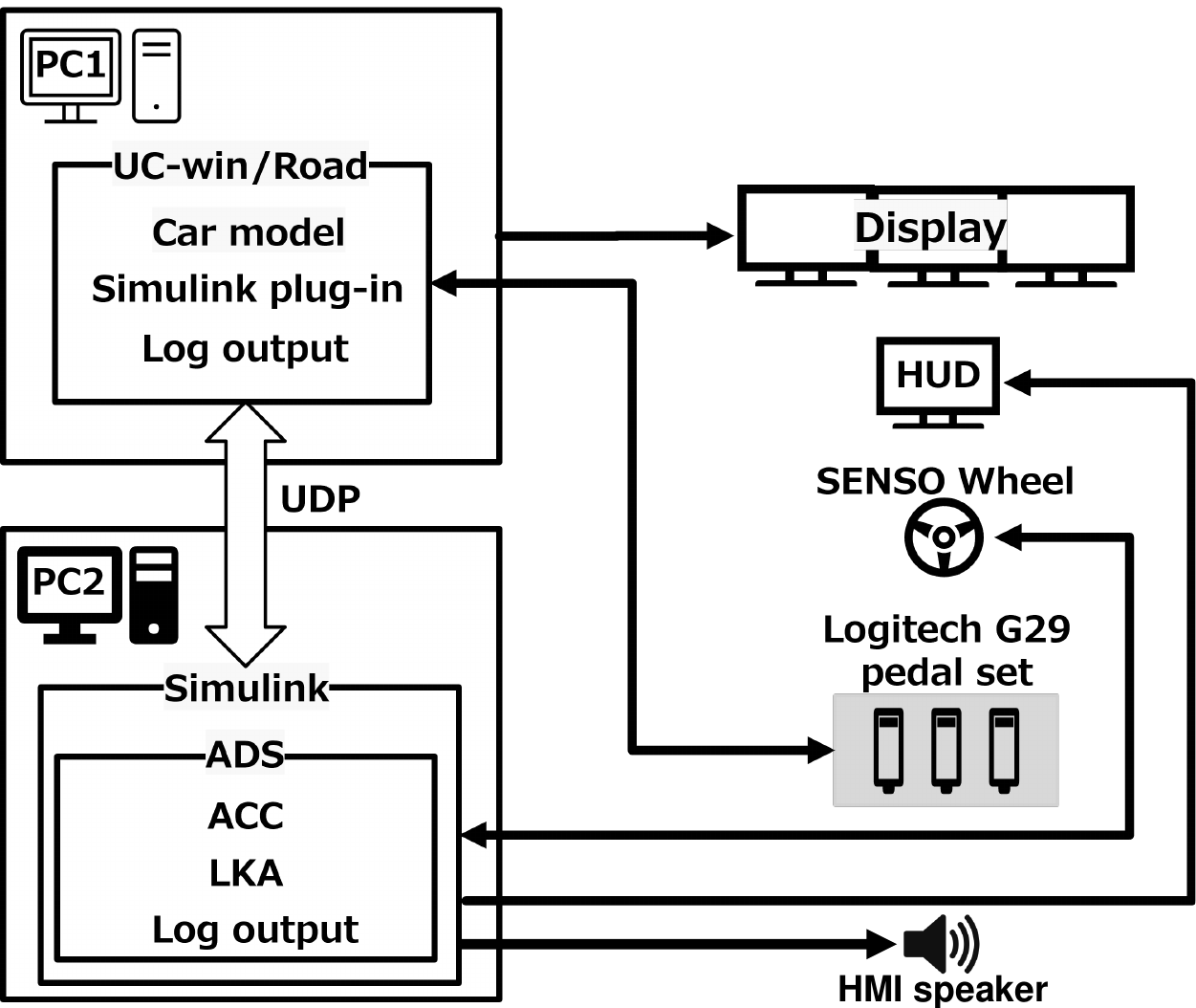}
\end{subfigure}
  \caption{Driving simulator used in this experiment with a HUD displaying ADS status and a speaker for RtI HMI voice cues.}
  \label{fig:DS}
  \vspace{-4mm}
\end{figure*}

\subsection{Reason Explanation of RtI Trigger}

Relying solely on pre-education may lead drivers to forget their knowledge of ADS. 
Therefore, continuous education of system limitation of ADS is considered effective.
This study proposes a voice-based method to provide information on trigger reasons in addition to the trigger cue.
As shown in Fig.~\ref{fig:trigger_reasons}, to prevent the driver from being hindered in responding quickly to the RtI due to receiving too much information, unlike in \cite{korber2018have}, the ADS explains the reason for the RtI via a voice cue when the driver reactivates the ADS after the take-over event.
The explanation voice cue started 5 seconds after reactivating the ADS.
If the visibility range is less than 40~m due to thick fog, the ADS will issue the voice cue: ``The reason for RtI is that it is difficult to recognize lanes and vehicles ahead in thick fog when it is impossible to see more than 40~m ahead.''  
Drivers who receive trigger reasons are expected to acquire an understanding of the ADS's system limitations and form an accurate mental model.

\section{Research Questions}

This study aims to verify the educational effectiveness of the proposed RtI HMI on drivers, such as improving their comprehension of the ADS's system limitations, and encouraging drivers to proactively take over control to ensure driving safety when the RtI system fails.
Additionally, considering that excessive prompts may increase the driver's workload, this study tries to verify the impact of the proposed RtI HMI on it.
Therefore, this study addresses the following four research questions:
\begin{description}
    \item[RQ~1:] Will the proposed \textit{RtI HMI w/ trigger cue \& reason} increase drivers' workload by providing more information prompts to them?
    \item[RQ~2:] Does the proposed \textit{RtI HMI w/ trigger cue \& reason} educate drivers to correctly understand the system limitations of the ADS?
    \item[RQ~3:] Can repeated use of the \textit{RtI HMI w/ trigger cue \& reason} be associated with earlier proactive takeovers by drivers before ADS deactivation and lower accident occurrence?
    \item[RQ~4:] Is better comprehension of the ADS limitations, \ie establishing a more accurate mental model of the ADS, associated with earlier proactive take-over behavior?
 \end{description}

\begin{figure*}[h!tb]

% \centering
  \begin{subfigure}{0.49\linewidth}
 %  \centering
  \includegraphics[width=0.7\linewidth]{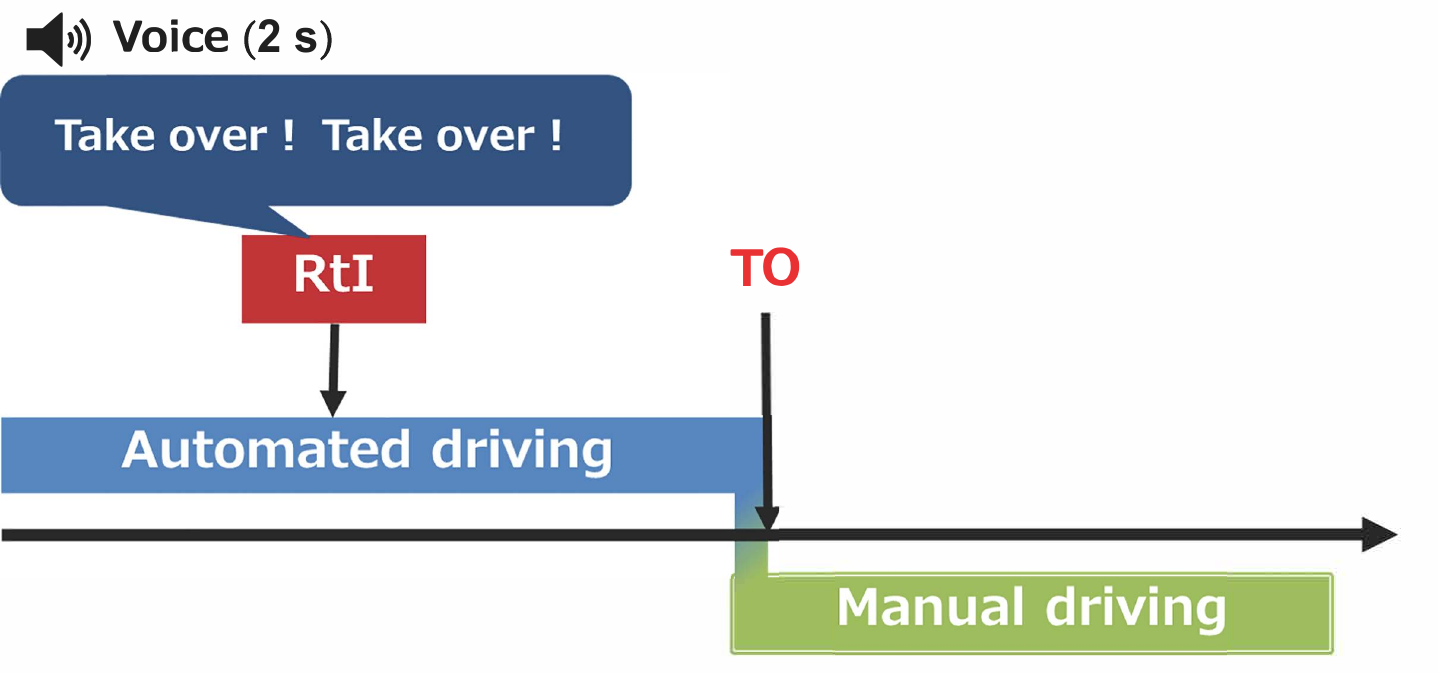}
  \caption{The RtI HMI \textit{w/o trigger cue} in thick fog.}
  \label{fig:baseline_fog}
  \end{subfigure}
%   \centering
  \begin{subfigure}{0.49\linewidth}
 %  \centering
  \includegraphics[width=0.7\linewidth]{images/Method/w_o_trigger_cue.pdf}
  \caption{The RtI HMI \textit{w/o trigger cue} in a sharp curve.}
  \label{fig:baseline_curve}
   \end{subfigure}
  %  \caption{The HMI of \textit{w/o trigger cue} that provides the driver information only the RtI in the experiment.}
  %\label{fig:baseline}
 \vspace{2mm}

 %\centering
  \begin{subfigure}{0.49\linewidth}
 %s\centering
  \includegraphics[width=0.7\linewidth]{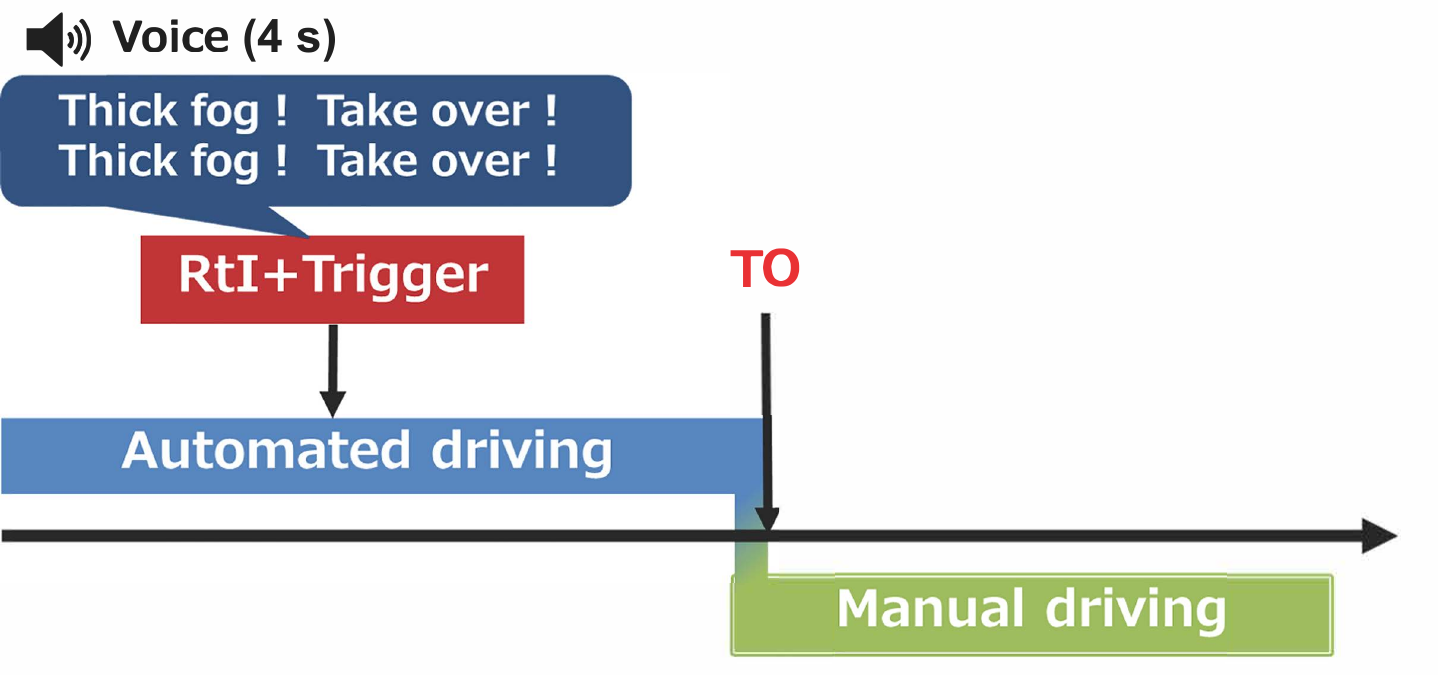}  
    \caption{The RtI HMI \textit{w/ trigger cue} in thick fog.}
     \label{fig:w_trigger_cues_fog}
  \end{subfigure}
  \begin{subfigure}{0.49\linewidth}
%   \centering
  \includegraphics[width=0.7\linewidth]{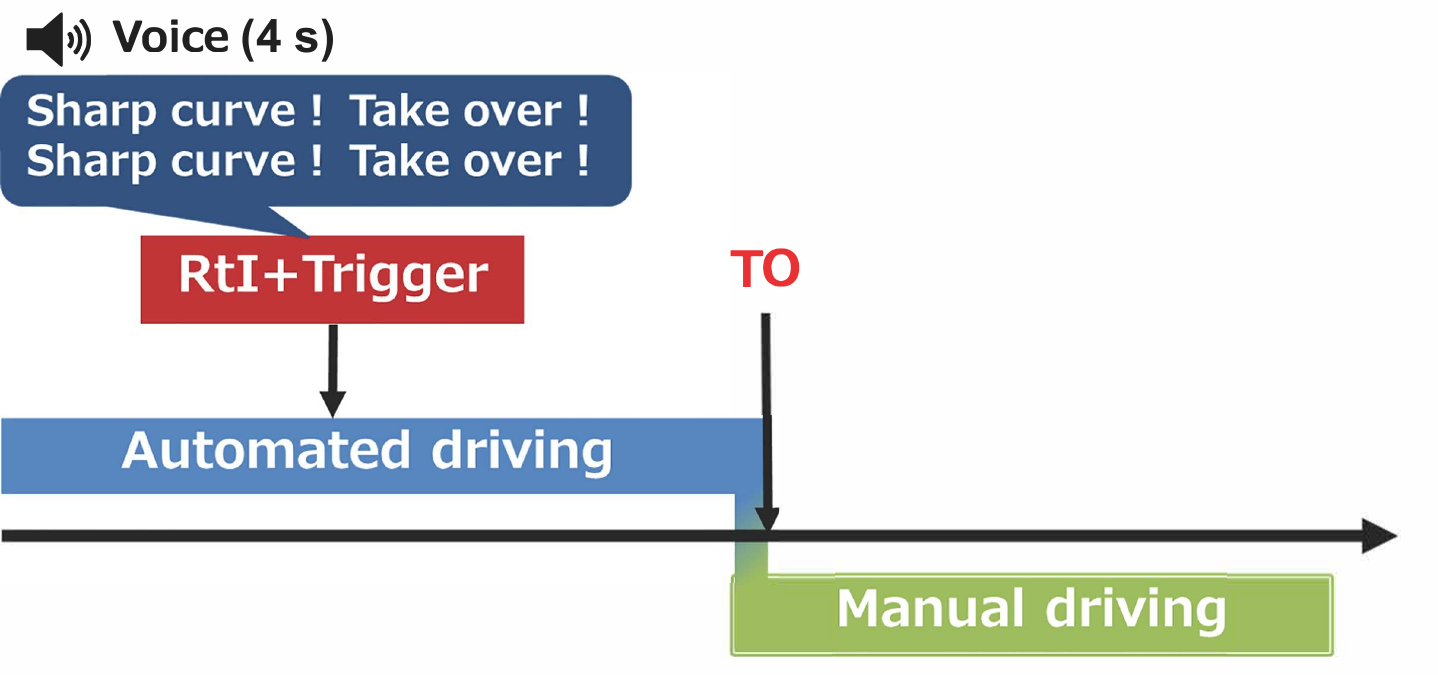}   
  \caption{The RtI HMI \textit{w/ trigger cue} in a sharp curve.}
   \label{fig:w_trigger_cues_curve}
  \end{subfigure}
  
  %\caption{Proposed HMI of \textit{w/ trigger cue} that provides the driver information about the trigger cues for the RtI in the experiment.}
  %\label{fig:w_trigger_cues}
\vspace{2mm}

 %\centering
  \begin{subfigure}{0.49\linewidth}
 % \centering
  \includegraphics[width=1\linewidth]{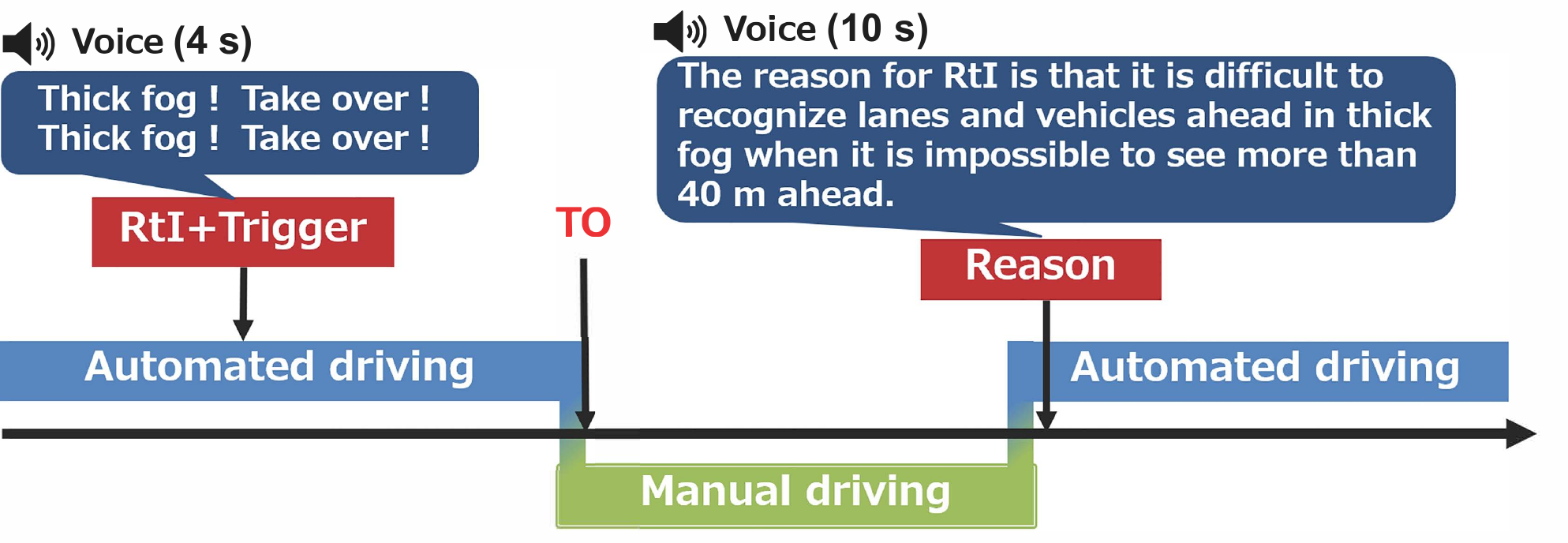}  
    \caption{The RtI HMI \textit{w/ trigger cue \& reason} in thick fog.}
    \label{fig:w_trigger_cues_reasons_fog}
  \end{subfigure}
  \begin{subfigure}{0.49\linewidth}
  % \centering
  \includegraphics[width=1\linewidth]{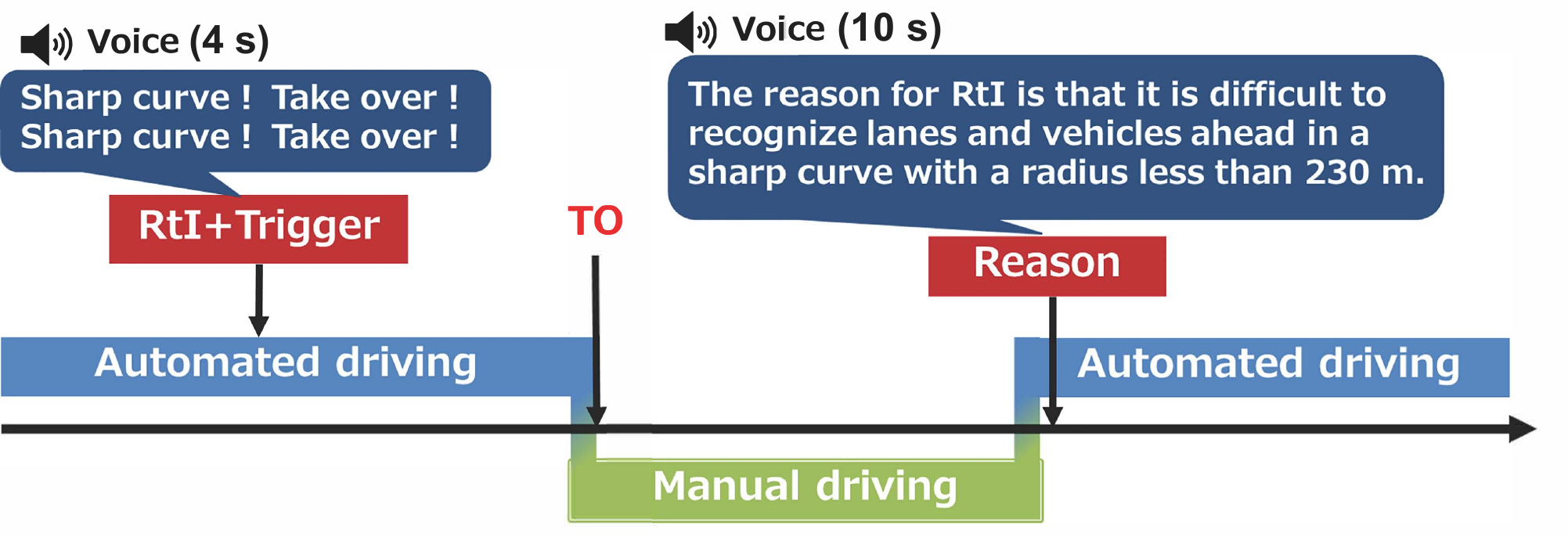}   
  \caption{The RtI HMI \textit{w/ trigger cue \& reason} in a sharp curve.}
  \label{fig:w_trigger_cues_reasons_curve}
  \end{subfigure}
  
  \caption{The three RtI HMIs used in the experiment for two RtI triggers (thick fog and sharp curve) in take-over~(TO) progress. All voice cues were presented in Japanese. The voice durations shown in the figures correspond to the Japanese utterances.}
  \label{fig:RtI_HMI}
\end{figure*}

\section{Experiment}
\label{sec:Experiment}

To answer the research questions, this study designed and conducted a between-group experimental study, using a driving simulator to validate the effectiveness of the proposed methods.
This experiment received approval from the Research Ethics Committee of Nara Institute of Science and Technology (No.~2022-I-56).

\subsection{Participants} \label{sec:participants}

This experiment recruited 45 participants (male:31, female: 14) aged 22 to 28 years (Avg.=23.36 years, Std.=1.42 years).
They were randomly assigned to three groups which were: the with trigger cue \& reason group (N=15), the with trigger cue group (N=15) and the without trigger cue group (N=15). 
All the participants had driving licenses in Japan and they had no experience using ADS.
Before the experiment, each participant provided informed consent and received a reward of 2,000 Japanese Yen in appreciation for the two hours of participation.

\subsection{Driving Simulator with a Level 3 ADS}
\label{sec:DS}

\subsubsection{Equipment}
The experiment used a driving simulator based on the UC-win/Road driving simulator by FORUM 8 Co., Ltd., as shown in Fig.~\ref{fig:DS}.
The vehicle operation hardware included a \textit{Logitech G29} pedal set, and a \textit{SensoWheel SD-LC} force feedback steering wheel. 
The driving simulator displayed graphics on three 55-inch LED displays with resolutions of $1920 \times1080$ pixels each (see Fig.~\ref{fig:DS}).

\subsubsection{Level 3 automated driving system (ADS)}

The driving simulator reproduces an automated vehicle (AV) which has a level~3 ADS including two main sub-systems: an adaptive cruise control (ACC) system and a lane-keeping assistance (LKA) system. 
The ACC system automatically regulates the accelerator and brake to follow the preceding vehicle.
Besides, it maintains a speed of 80~km/h if there is no preceding vehicle.
The driver can override the ACC function during automated driving by pressing the accelerator pedal.
The LKA system automatically controls the steering wheel to keep the vehicle centered within the lane by detecting the lane markings. 
The driver can activate the ACC and LKA systems by pressing the clutch pedal during manual driving. 
In addition, the driver can deactivate the ACC and LKA systems by pressing the brake pedal or turning the steering wheel with a torque greater than 5~N$\cdot$m.
As shown in Fig.~\ref{fig:DS}, the state of ADS, i.e., ON or OFF, is shown on a head-up display (HUD).
In particular, the green and gray icons indicate that the ADS is ON and OFF, respectively.

\subsubsection{System limitations}
\label{sec:system_limitations}

According to Article~15 (Curve Radius) of the Road Structure Ordinance~\cite{JP_RSO}, the minimum curve radius for a design speed of 80~km/h is 230~m.
In addition, \citep{ zhou2021does, chaabani2018estimating, pereira2024weather} defined thick fog as conditions where visibility is below 40~m, and \citet{zhou2021does} used it as the ODD limitation in their Level~3 ADS takeover experiment.
Therefore, this study incorporated two specific system limitations, \ie the conditions under which the RtI was issued:
\begin{itemize}
    \item[(1)] curve with a radius of less than 230~m~\cite{JP_RSO},
    \item[(2)] visibility range less than 40~m~\cite{zhou2021does}.
\end{itemize}
The RtI is triggered when any of the conditions exceed their thresholds.
Note that the ADS can function normally even in thin fog if the visual range remains above 40 m. 
In such cases, the RtI will not be triggered.

\subsection{Conditions of RtI HMI}
Based on the two conditions in our pre-study~\cite{Matsuo2024ICHMS}, the three conditions of the RtI HMI were designed and used in the experiment, as shown in Fig.~\ref{fig:RtI_HMI}. 
Note that all voice cues from the three RtI HMIs were in Japanese.
These voices were generated from text using Text-To-Speech application called \textit{Microsoft Nanami Online} and prerecorded.

\subsubsection{RtI HMI \textit{w/o trigger cue}}
The baseline method shown in Figs.~\ref{fig:baseline_fog} and \ref{fig:baseline_curve} used a conventional voice-based RtI HMI that issues a voice cue in 2 seconds of ``Take over! Take over!'' without the trigger cue and the reason explanation.
This group was named \textit{w/o trigger cue} group.

\subsubsection{RtI HMI \textit{w/ trigger cue}}
Figs.~\ref{fig:w_trigger_cues_fog} and \ref{fig:w_trigger_cues_curve} show the RtI HMI used in \textit{w/ trigger cue} group that provides only trigger cues.
The ADS issues the RtI with their corresponding voice cues from in-car speakers under the following conditions according to the system limitations:
1) If a curve with a radius less than 230~m is detected, the ADS will issue the RtI with the voice cue in 4 seconds: ``Sharp curve, take over! Sharp curve, take-over!''.
2) If the visibility range is less than 40~m due to thick fog, the ADS will issue the RtI with the voice cue in 4 seconds: ``Thick fog, take over! Thick fog, take over!''.

\subsubsection{RtI HMI \textit{w/ trigger cue \& reason}}
Figs.~\ref{fig:w_trigger_cues_reasons_fog} and ~\ref{fig:w_trigger_cues_reasons_curve} show the RtI HMI used in \textit{w/ trigger cue \& reason} group that provides detailed information regarding the trigger cues and post hoc explanations of the reason.
The ADS issues the RtI with their corresponding voice cues from in-car speakers under the following conditions according to the system limitations: 
1) If a curve with a radius less than 230~m is detected, the ADS will issue the RtI with the voice cue in 4 seconds: ``Sharp curve, take over! Sharp curve, take-over!''.
After the driver reactivates the ADS after the take-over event, it will issue the voice cue for the reason explanation in 10 seconds: ``The reason for RtI is that it is difficult to recognize lanes and vehicles ahead in a sharp curve with a radius less than 230~m.''
2) If the visibility range is less than 40~m due to thick fog, the ADS will issue the RtI with the voice cue  in 4 seconds: ``Thick fog, take over! Thick fog, take over!''.
After the driver reactivates the ADS after the take-over event, it will issue the voice cue for the reason explanation in 10 seconds: ``The reason for RtI is that it is difficult to recognize lanes and vehicles ahead in thick fog when it is impossible to see more than 40~m ahead.''

\subsection{Scenarios}

\begin{figure*}[h!tb]
 \centering
  \begin{subfigure}[b]{0.49\linewidth}
%  \centering
  \includegraphics[width=0.9\linewidth,clip,trim=0 5 0 0]{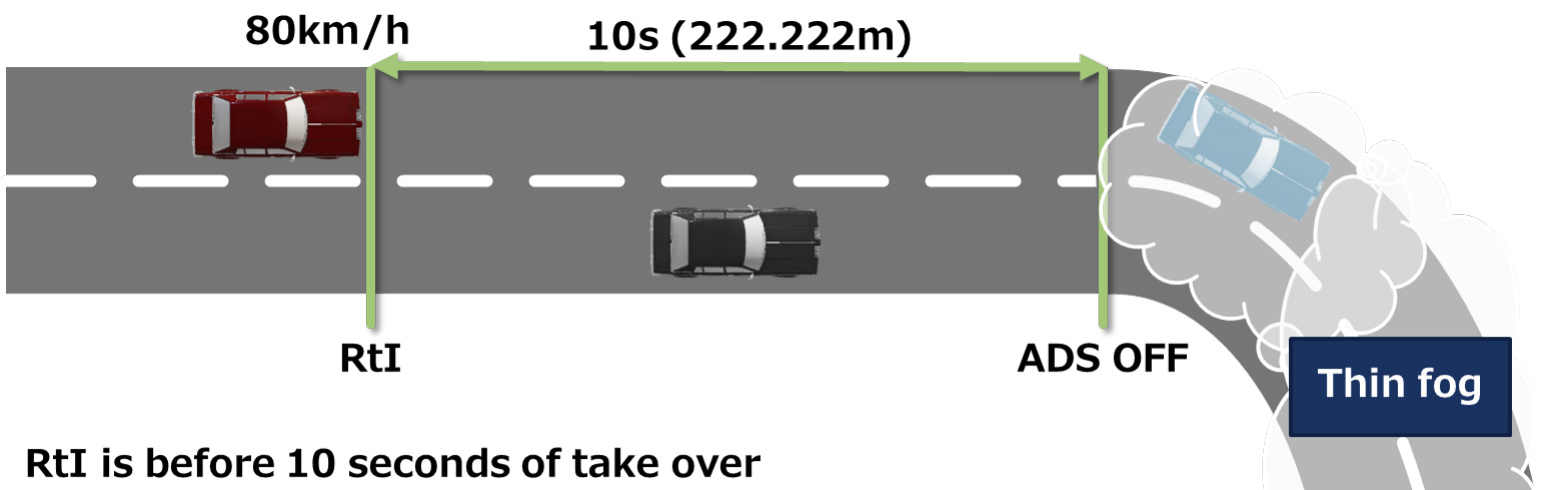}  
    \caption{A sharp curve road with thin fog.}
  \end{subfigure}
  \begin{subfigure}[b]{0.49\linewidth}
 %  \centering
  \includegraphics[width=1\linewidth]{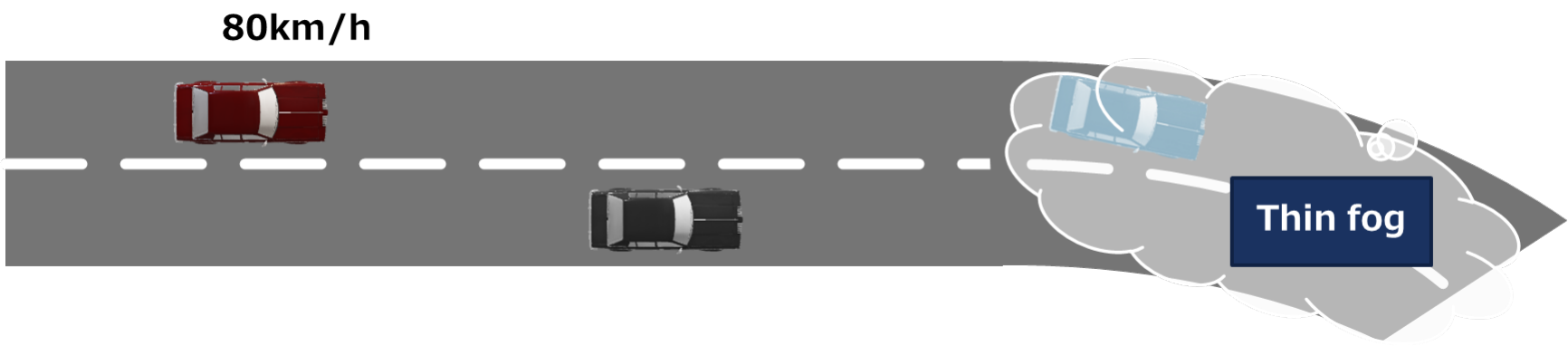}   
  \caption{A slight curve road with thin fog.}
  \end{subfigure}\\
 % \vspace{2mm}
  \begin{subfigure}[b]{0.49\linewidth}
% \centering
  \includegraphics[width=0.95\linewidth]{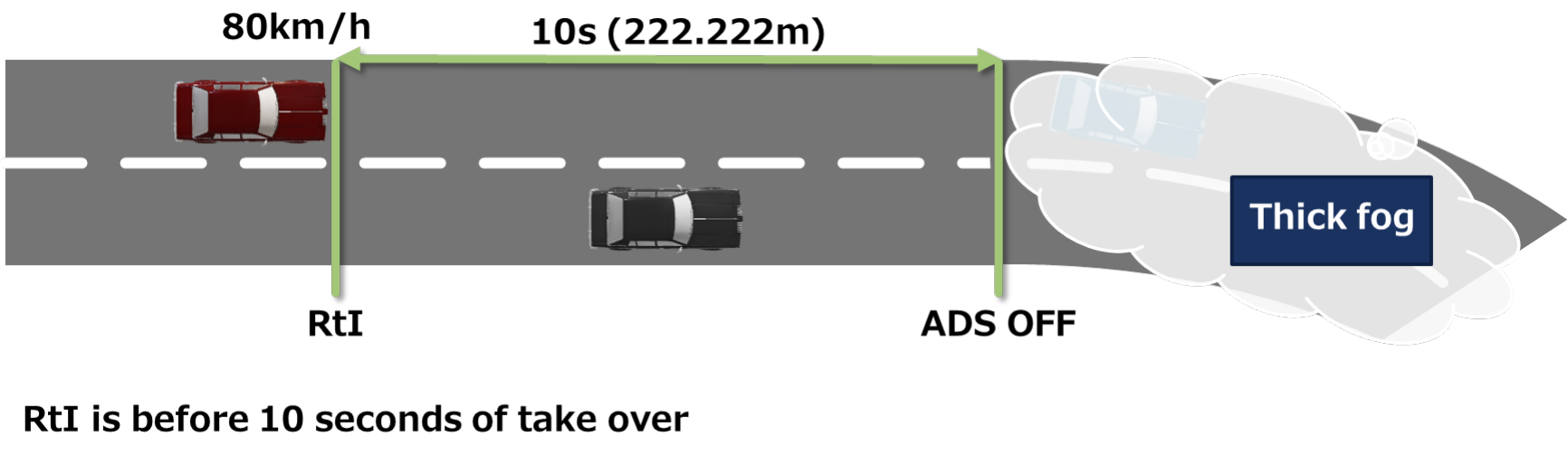}   
  \caption{A slight curve road with thick fog.}
  \end{subfigure}
   \begin{subfigure}[b]{0.49\linewidth}
%  \centering
  \includegraphics[width=0.9\linewidth,clip,trim=0 10 0 0]{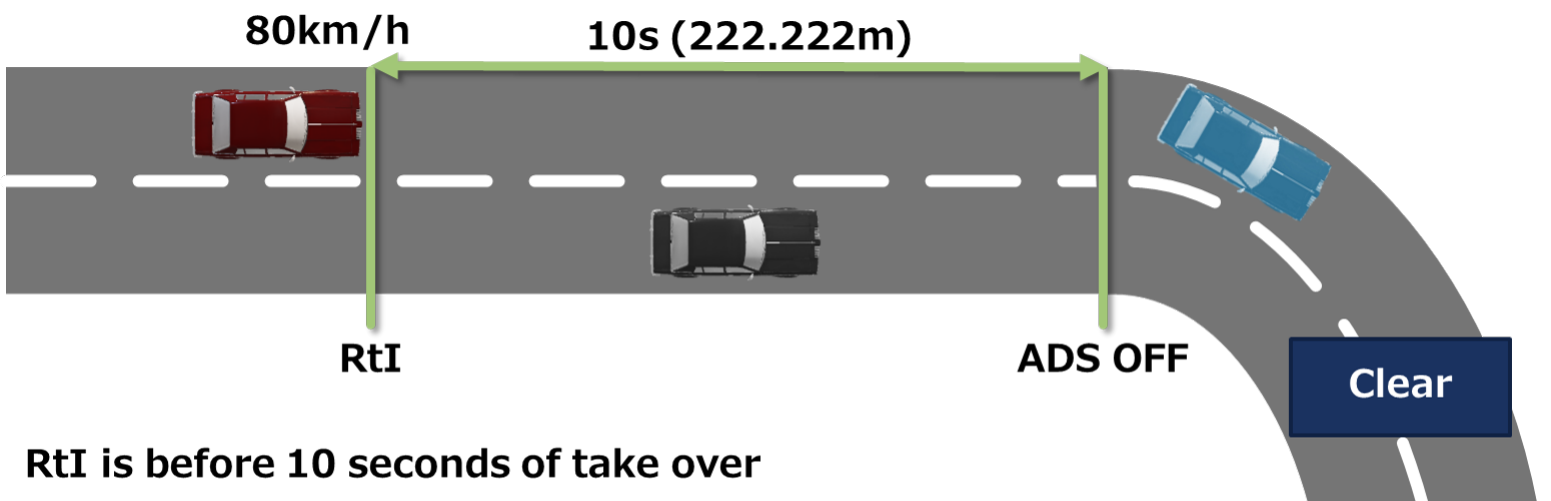}    
  \caption{A sharp curve road without fog.}
  \end{subfigure}
  \\
 % \vspace{2mm}
  \begin{subfigure}[b]{0.49\linewidth}
 %  \centering
  \includegraphics[width=1\linewidth]{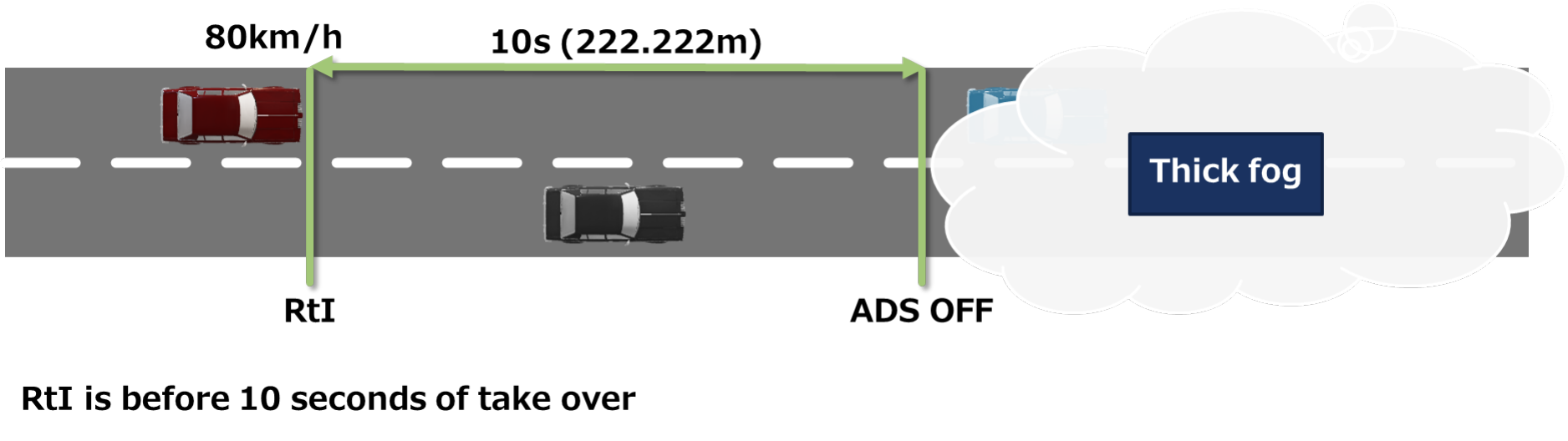}
  \caption{A straight road with thick fog.}
  \end{subfigure}
  \begin{subfigure}[b]{0.49\linewidth}
%   \centering
  \includegraphics[width=1\linewidth]{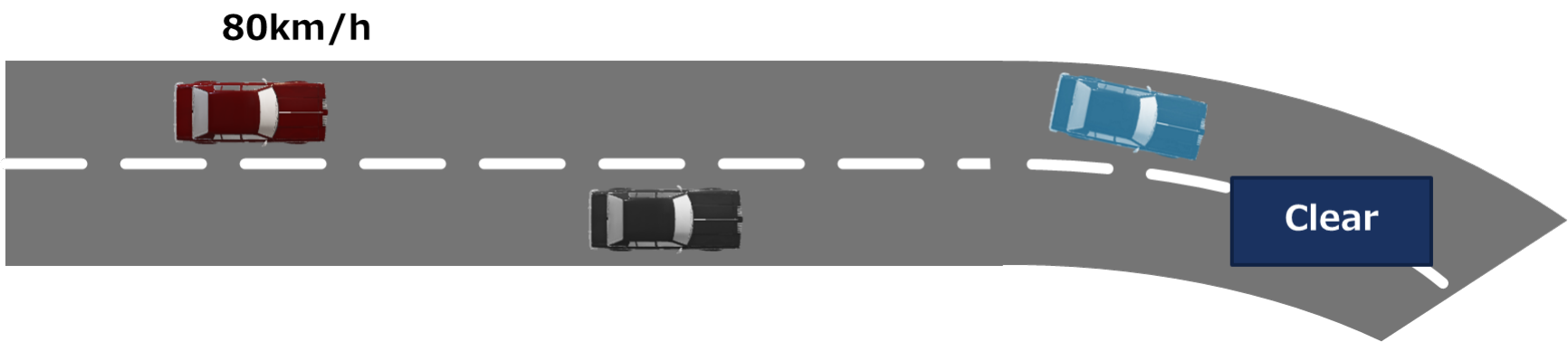}   
  \caption{A slight curve road without fog.}
  \end{subfigure}
  \\
%  \vspace{2mm}
  \begin{subfigure}[b]{0.49\linewidth}
   %\centering
  \includegraphics[width=1\linewidth]{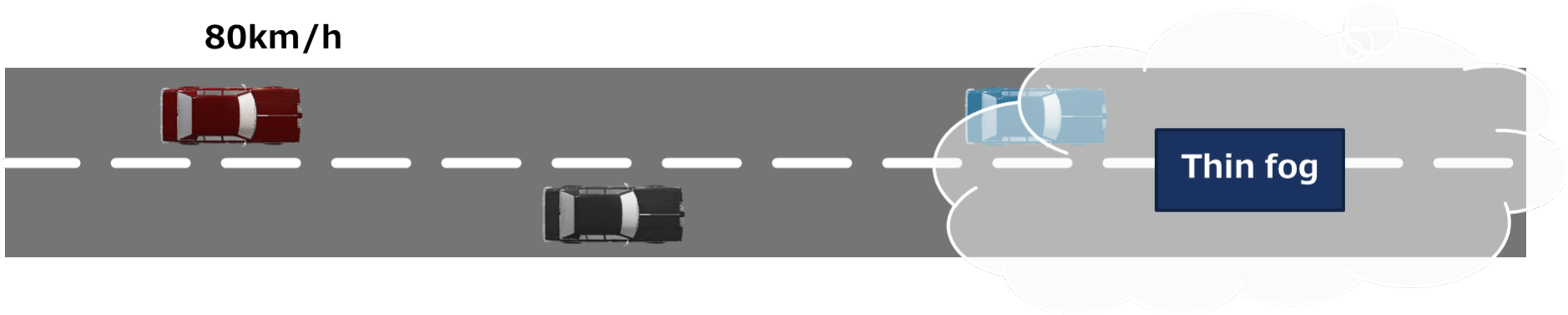}   
  \caption{A straight road with thin fog.}
  \end{subfigure}
  \begin{subfigure}[b]{0.49\linewidth}
  % \centering
  \includegraphics[width=0.9\linewidth,clip,trim=0 10 0 0]{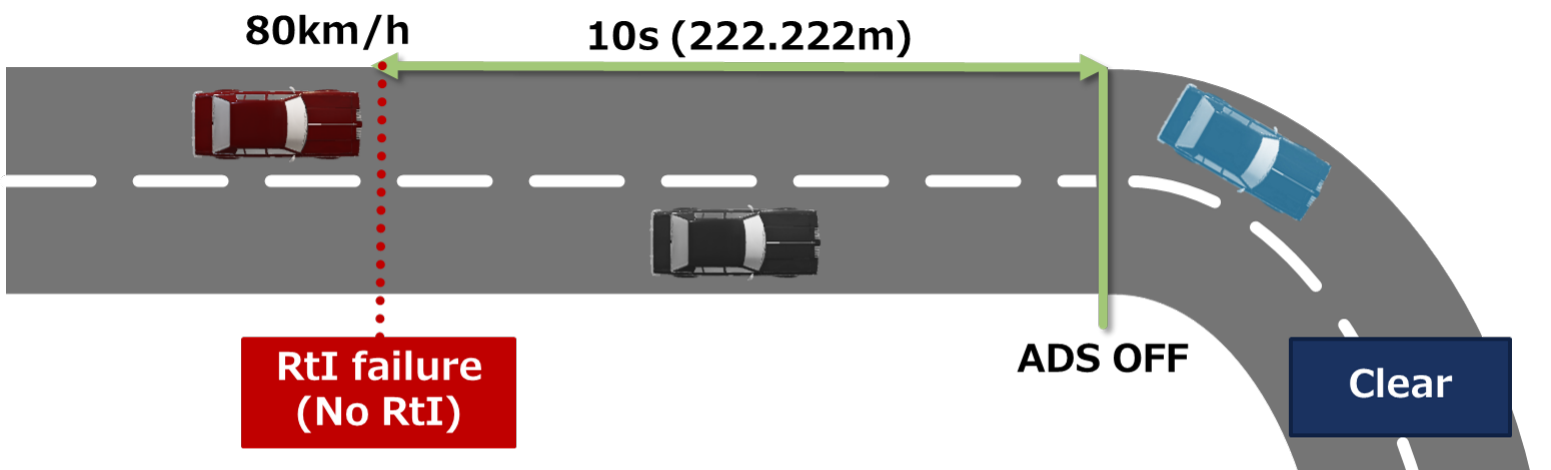}   
  \caption{A sharp curve road without fog but the RtI is failed.}
  \end{subfigure}
    \caption{The various traffic scenarios used in the experiment, where the red car is the ego AV with a speed of 80 km/h. The scenarios (a), (b), and (c) are used in the learning phase I; (d) and (e) are  used in the learning phase II; and (f) and (g) are used in the learning phase III; (h) is used in the test phase.}
  \label{fig:scenario}
  %\vspace{-2mm}
\end{figure*}

\begin{table*}[h!tb]
\footnotesize
\setlength\tabcolsep{6pt}
  \caption{Experiment scenarios (conditions for RtI to be issued: a curve with R $<$ 230~m or the visibility range $<$ 40~m)}  
  \label{table:Scenario}
  \centering
  \scalebox{1}[1]{
  \begin{tabular}[H]{lccrcrr}
    \toprule
    \multicolumn{1}{c}{Phases}&\begin{tabular}[c]{@{}c@{}}Trial\\ No.\end{tabular}  & \begin{tabular}[c]{@{}c@{}}Scenario\\ shown in Fig.~\ref{fig:scenario}\end{tabular}    & \multicolumn{1}{c}{Visibility conditions} & Road conditions  &  \multicolumn{1}{c}{RtI}  &  \multicolumn{1}{c}{Trigger of RtI} \\
   \midrule
   Learning phase I & 1 & (a)  &  Thin fog (Visibility range $\geq$ 40 m)  & Sharp curve (R=180~m)  & Issued  & Sharp curve \\
   Learning phase I & 2 & (b)  &  Thin fog (Visibility range $\geq$ 40 m)  & Slight curve (R=280~m)  & Not issued  & None \\
   Learning phase I & 3 & (c)  &  Thick fog (Visibility range $<$ 40 m)  & Slight curve (R=290~m)  & Issued  & Thick fog \\
   Learning phase I & 4 & (b)  &  Thin fog (Visibility range $\geq$ 40 m)  & Slight curve (R=300~m)  & Not issued  & None \\
   Learning phase I & 5 & (c)  &  Thick fog (Visibility range $<$ 40 m)  & Slight curve (R=310~m)  & Issued  & Thick fog \\
   Learning phase I & 6 & (a)  &  Thin fog (Visibility range $\geq$ 40 m)  & Sharp curve (R=170~m)  & Issued  & Sharp curve \\
   Learning phase I & 7 & (c)  &  Thick fog (Visibility range $<$ 40 m)  & Slight curve (R=320~m)  & Issued  & Thick fog \\
   Learning phase I & 8 & (b)  &  Thin fog (Visibility range $\geq$ 40 m)  & Slight curve (R=330~m)  & Not issued  & None \\
   Learning phase I & 9 & (a)  &  Thin fog (Visibility range $\geq$ 40 m)  & Sharp curve (R=160~m)  & Issued  & Sharp curve \\ %\midrule
   Learning phase II & 10 & (d)  &  Clear (Visibility range $\geq$ 40 m) & Sharp curve (R=150~m)  & Issued  & Sharp curve \\
   Learning phase II & 11 & (e)  &  Thick fog (Visibility range $<$ 40 m)  & Straight road  & Issued  & Thick fog \\%\midrule
   Learning phase III & 12 & (f)  &  Clear (Visibility range $\geq$ 40 m) & Slight curve (R=330~m)  & Not issued  & None \\ 
   Learning phase III &  13 & (g)  &  Thin fog (Visibility range $\geq$ 40 m)  & Straight road  & Not issued  & None \\ \midrule
    Test phase & 14  & (h)  &  Clear (Visibility range $\geq$ 40 m) & Sharp curve (R=150~m)  & \begin{tabular}[r]{@{}r@{}}ODD is exceeded\\ Not issued\\(RtI failure)\end{tabular}  & Sharp curve \\ 
    \bottomrule
  \end{tabular}
  }
  \vspace{-2mm}
\end{table*}

\begin{figure*}[h!tb]
 \centering
  \begin{subfigure}[b]{0.245\linewidth}
  \centering
  \includegraphics[width=1\linewidth]{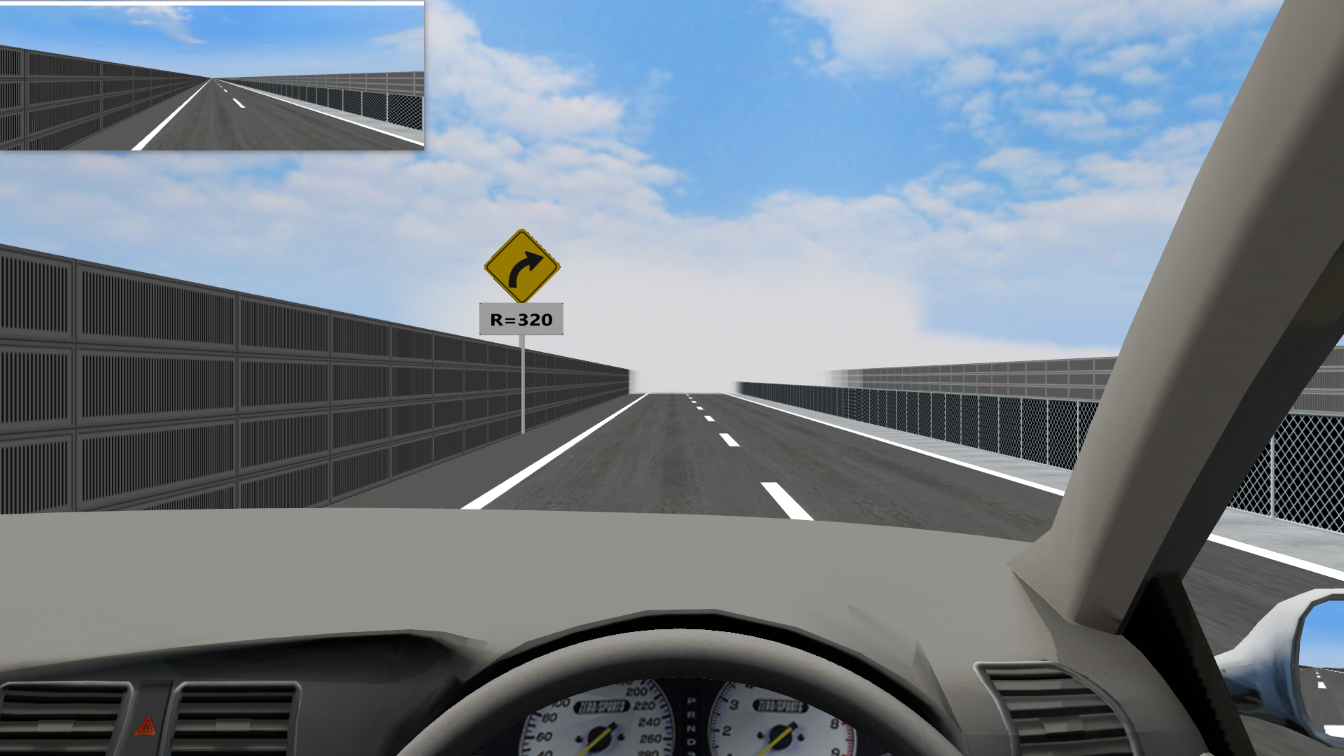}  
    \caption{Slight curve with thick fog.\\ \centering(Trial No.7)}
  \end{subfigure}
  \begin{subfigure}[b]{0.245\linewidth}
   \centering
  \includegraphics[width=1\linewidth]{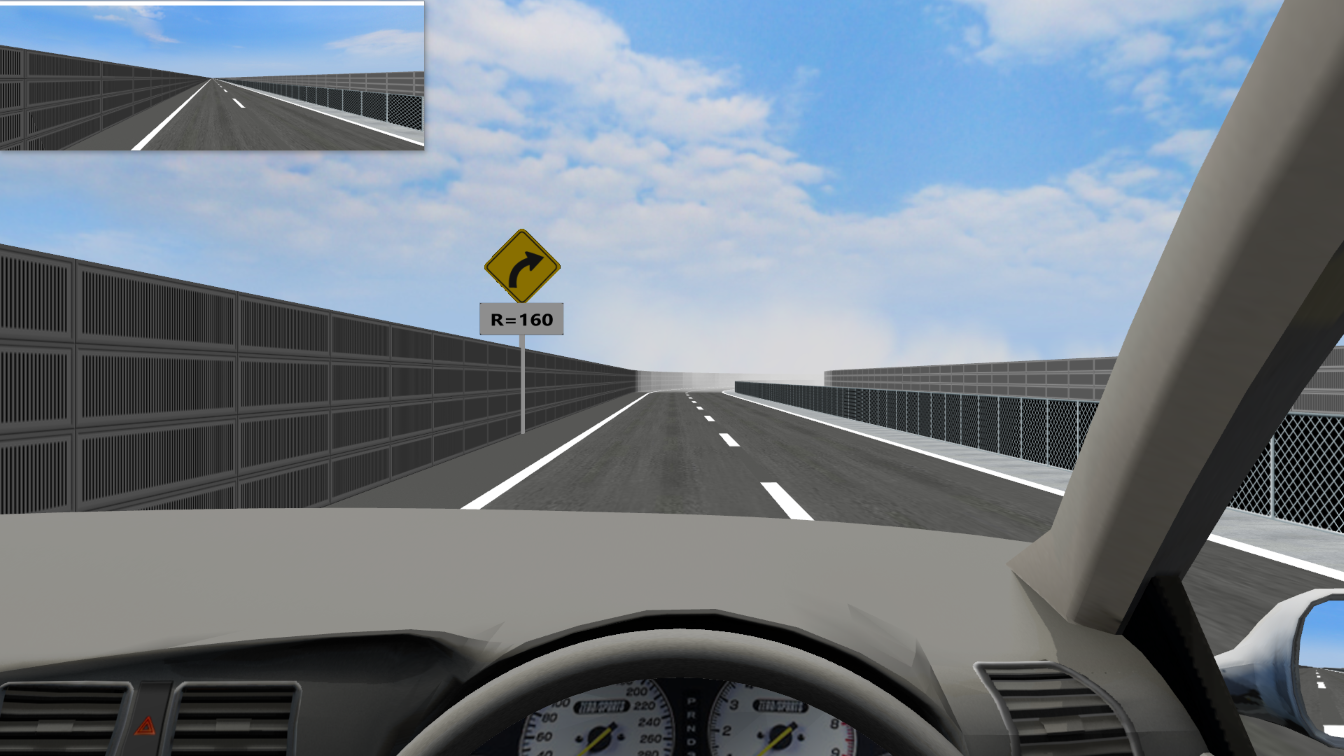}   
  \caption{Sharp curve with thin fog.\\ \centering(Trial No.9)}
  \end{subfigure}
  \begin{subfigure}[b]{0.245\linewidth}
 \centering
  \includegraphics[width=1\linewidth]{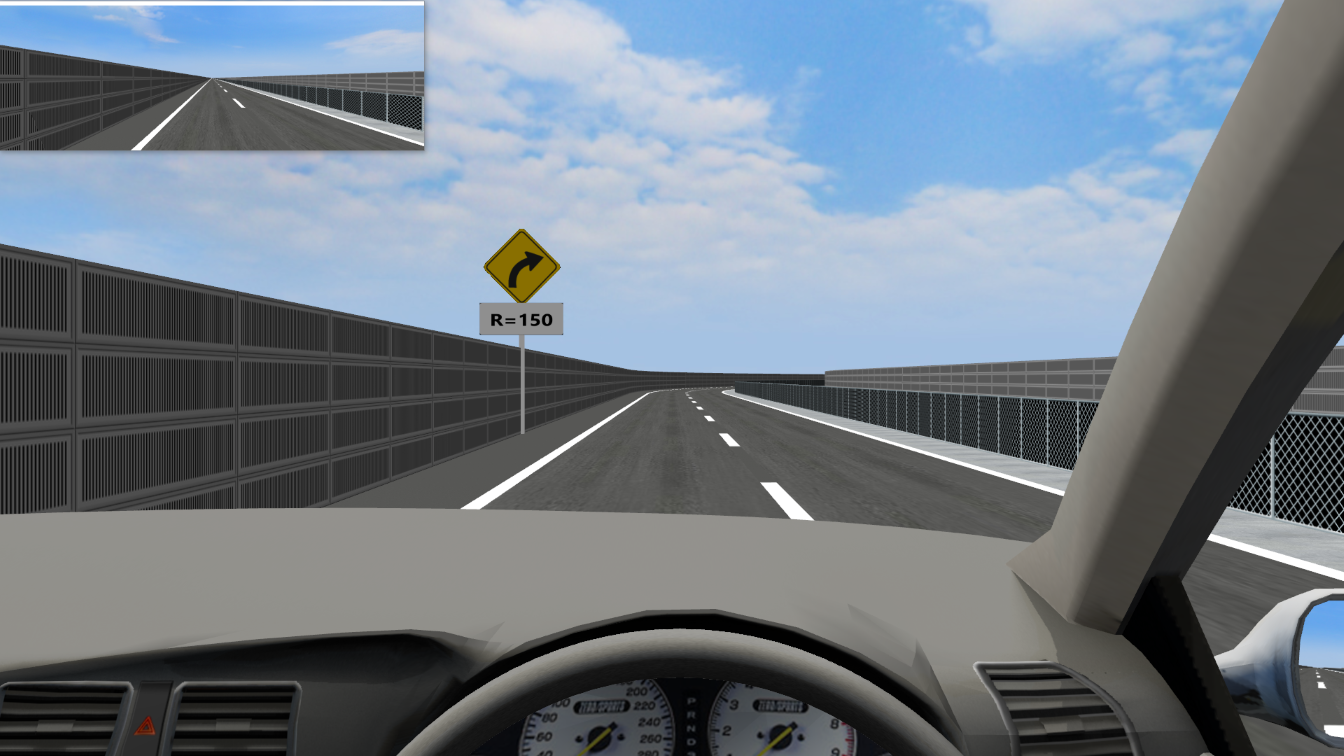}   
  \caption{Sharp curve without fog.\\  \centering(Trial No.10)}
  \end{subfigure}
   \begin{subfigure}[b]{0.245\linewidth}
  \centering
  \includegraphics[width=1\linewidth]{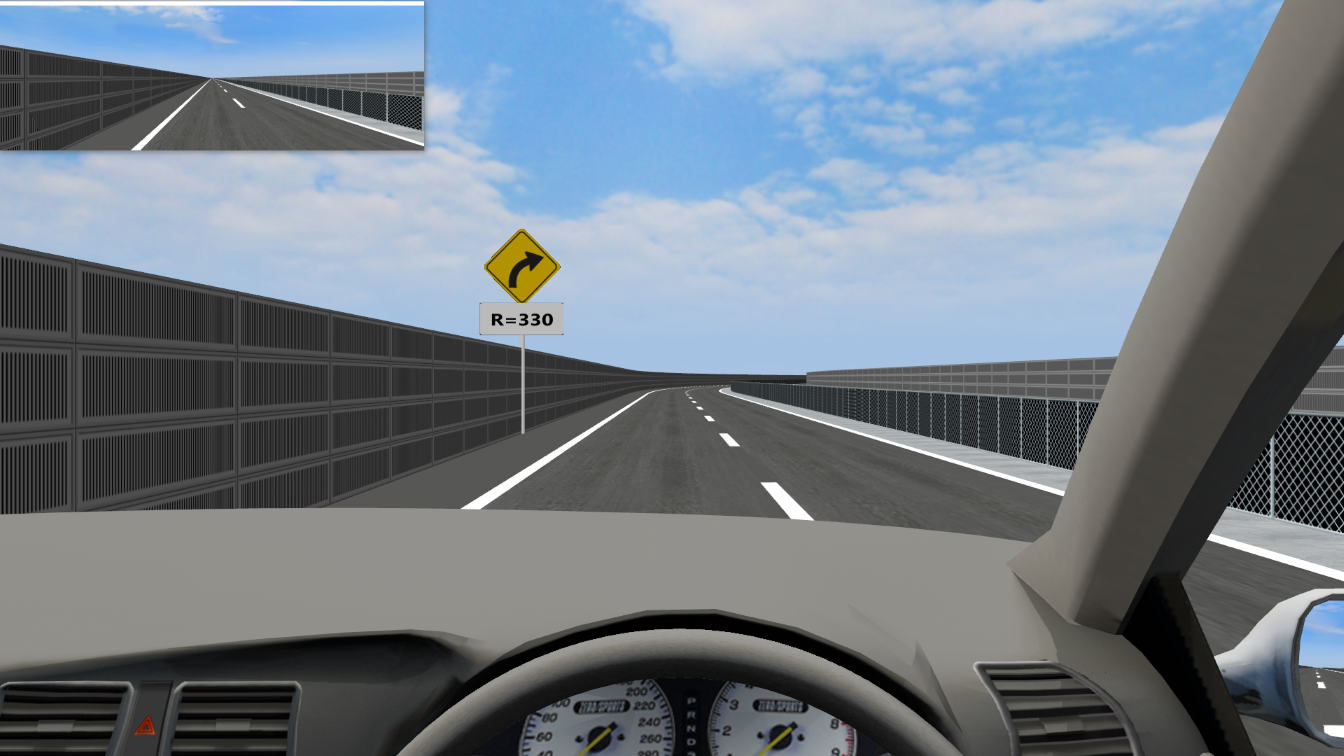}    
  \caption{Slight curve without fog.\\ \centering(Trial No.12)}
  \end{subfigure}
    \caption{Examples of the designed traffic scenarios as seen from the driver's view. (a) and (b) show thick and thin fog scenarios, respectively, while (c) and (d) show sharp and slight curves. Traffic signs show the curvature of curve roads.}
  \label{fig:scenario_from driver}
  \vspace{-4mm}
\end{figure*}

This study focuses on situations in which there appear to be multiple potential triggers contributing to RtI as perceived by drivers. 
Specifically, this study is honing in on a specific scenario involving the simultaneous presence of fog and curves.
The reason for using these two factors is that in practical scenarios, the curvature information of the curve can be obtained in advance from the geographical information in the navigation system. 
Additionally, the density of fog can be recognized by camera~\cite{mori2007fog,cao2023fog} and millimeter wave radar~\cite{mori2007fog}.

As shown in Fig.~\ref{fig:scenario} and detailed in Table~\ref{table:Scenario}, this study have meticulously crafted eight distinct driving scenarios tailored for employment on the two-lane highway. 
These particular scenarios have been meticulously selected to serve as the foundation for the four phases of the experiment.
To elaborate, scenarios (a), (b), and (c) are used in the learning phase I, which appears to encompass multiple potential RtI triggers from the driver's perspective.
However, the ADS issues an RtI by only one correct RtI trigger in (a) and (c), respectively.
Moreover, the ADS did not issue RtI in scenario (b) because the visibility and road conditions did not exceed its system limitations.
The experiment repeated these three scenarios three times to train the drivers to understand the system limitations of the ADS accurately.
In particular, when using the proposed RtI HMI to cue the correct RtI trigger to the driver, this study assumes that the driver can develop a better understanding of the system limitations by comparing multiple potential triggers with the correct trigger.
Table~\ref{table:Scenario} shows details of the visibility conditions and road conditions and the correct triggers for each of the nine scenarios (scenarios (a), (b), and (c) repeated three times).

The learning phase II comprises scenarios (d) and (e), each having only one potential RtI trigger and both appropriately trigger the RtI.
During Learning Phase III, scenarios (f) and (g) each involve only one potential trigger. However, since these triggers remain within the system's ODD and their visibility and road conditions do not exceed system limits, they do not activate an ADS RtI (see Table~\ref{table:Scenario}).

Finally, to test whether participants correctly understood the system limitations of the ADS, a test phase was designed based on scenario (h) to simulate an RtI failure scenario. 
In this scenario, the ADS should issue an RtI because the ADS will exceed the system limitation.
Unfortunately, as a result of a system failure, the driver is unable to receive pertinent information regarding the RtI.
When the ADS has been exceeded its system limitations, then it is forced to turn off.
Thus, if the driver fails to take over within approximately 1.5 seconds after the ADS turned off, the vehicle will collide with the guardrail at the sharp curve.
For this test phase, we considered that if drivers were difficult to establish a correct mental model of the ADS, they are likely to over-trust and over-reliance on the system, as described in~\cite{liu2019driving}.
Such over-trust and over-reliance might prevent drivers from detecting system issues in a timely manner, potentially leading to delayed interventions and even accidents.
Therefore, in such a critical situation, only drivers who have a correct understanding of the limitations of the ADS will notice that something is wrong and will take proactive action to take over vehicle control and avoid a collision. 
This highlights the importance of drivers being well-informed about the system's capabilities and limitations, allowing them to respond effectively to unforeseen circumstances.

Fig.~\ref{fig:scenario_from driver} shows some examples of the designed traffic scenarios with different visibility (thick fog and thin fog) and road conditions (sharp curve and slight curve) as seen from the driver's view. 
It is important to note that there is a traffic sign indicating the curvature of the curve ahead based on the conditions of the Japan transportation system.
In addition, the surrounding vehicles were placed in front of the ego vehicle and the right lane to enhance the realism of the nine scenarios above. 
These surrounding vehicles maintain a safe following distance (more than 50~m) from the ego vehicle and travel at 80--100 km/h.
These vehicles do not suddenly decelerate or proactively approach the experimental vehicle, posing a risk to the vehicle.
The ego vehicle was setting the target speed of ACC to 80 km/h, and the maximum speed of the ego vehicle was limited to 110 km/h.

\subsection{Procedure}

First, participants were informed that the driving simulator was simulating a level 3 AV and were provided with the definition of SAE level 3 ADS.
Next, they were introduced to the equipment and operation of the simulator, with a particular focus on how to take control of the vehicle when using the ADS.
Subsequently, the participants were conducted a pre-instruction regarding the system limitations of the ADS using an instruction manual.
This manual describes various conditions of the system limitations, such as the set conditions of visibility and road curvature with their respective thresholds (see Section~\ref{sec:system_limitations}).
Additionally, other additional conditions of system limitation that did not occur in this experiment were also listed in the instruction manual to avoid making the set conditions conspicuous.
These additional conditions were taken from the instruction manuals of vehicles with ACC and LKA currently available on the market, \eg road gradients, icy roads, tunnel entrances, and exits.
After the pre-instruction, the participants were asked to practice using a driving simulator, such as manual driving, turning the ADS ON / OFF, and take-over.

The main experiment was conducted after the participants fully understood how to use the driving simulator.
Each participant experienced the 14 driving scenarios listed in Table~\ref{table:Scenario}, and no compulsory non-driving related task (NDRT) was requested during automated driving.
Participants were required to respond to the RtI when it was issued.
After successfully taking over control, participants were instructed to resume the ADS at a time they deemed appropriate.
In particular, the duration of manual driving after the take-over was not restricted and was left to the discretion of the participants.

At the end of the driving experiment, the NASA-TLX questionnaire~\cite{hart1988development} and a comprehension test regarding the ADS system limitations (see Section~\ref{sec:post-test}) were conducted.

\subsection{Measurements}
The following four measurements were obtained during the experiment to answer the four research questions described in Section~\ref{sec:RQ}.

\subsubsection{Weighted workload score measured via NASA-TLX}
\label{sec:NASA-TLX}

To confirm whether the proposed RtI HMI increased workload, participants completed the Japanese version of the NASA-TLX~\cite{haga1996japanese} after all driving scenarios.
We have further clarified in the Measurements section that the measured workload was assessed using the weighted workload score (WWL$_{\text{score}}$) of the NASA-TLX, which represents an overall workload index calculated by weighting and combining the sub-scales of mental demand, physical demand, temporal demand, performance, effort, and frustration~\cite{haga1996japanese}.

\subsubsection{Post-experiment comprehension test}
\label{sec:post-test}

To check the educational effectiveness of the
participants on the ADS's system limitations after using the proposed RtI HMI multiple times, they were asked to complete a comprehension test after answering NASA-TLX.
As shown in Table~\ref{table:Post-experiment comprehension test}, this comprehension test presented a list of 13 scenarios of the learning phase (excluding the test phase, \ie RtI failure scenario) from Table~\ref{table:Scenario}, along with their corresponding visibilities and road conditions specified by specific values.
The participants were tasked with determining whether the ADS could perform automated driving tasks under the given conditions by three options: ``Yes,'' ``No'' or ``I don't know.''
One point was awarded for each correct answer, while no points were given for incorrect answers or selecting ``I don't know.''
Overall, the comprehension test results were scored from 0 to 13 points.

\begin{table}[!t]
\footnotesize
\renewcommand{\arraystretch}{1}
\setlength\tabcolsep{3pt}
  \caption{Post-experiment comprehension test sheet with a question: ``Do you think the ADS you experienced can perform automated driving under the following conditions?'' The correct answers are shown by \donebox.}  
  \label{table:Post-experiment comprehension test}
  \centering
  \begin{tabular}[H]{cccccc}
    \toprule
    No. & \begin{tabular}[c]{@{}c@{}}Visibility conditions \end{tabular} & \begin{tabular}[c]{@{}c@{}}Road conditions \end{tabular} & Yes & I don't know & No \\
   \midrule
    1 & \begin{tabular}[c]{@{}c@{}}Thin fog \\(Visibility range $\geq$ 40 m) \end{tabular} & \begin{tabular}[c]{@{}c@{}}Sharp curve \\(R=180~m) \end{tabular} & \nonbox  & \nonbox & \donebox \\   \midrule
    2 & \begin{tabular}[c]{@{}c@{}}Thin fog \\(Visibility range $\geq$ 40 m) \end{tabular} & \begin{tabular}[c]{@{}c@{}}Slight curve \\(R=280~m) \end{tabular} & \donebox & \nonbox &\nonbox \\   \midrule
    3 & \begin{tabular}[c]{@{}c@{}}Thick fog \\(Visibility range $<$ 40 m) \end{tabular} & \begin{tabular}[c]{@{}c@{}}Slight curve \\(R=290~m) \end{tabular} & \nonbox  & \nonbox & \donebox \\   \midrule
      4 & \begin{tabular}[c]{@{}c@{}}Thin fog \\(Visibility range $\geq$ 40 m) \end{tabular}  & \begin{tabular}[c]{@{}c@{}}Slight curve \\(R=300~m) \end{tabular} & \donebox  & \nonbox & \nonbox \\   \midrule
    5 & \begin{tabular}[c]{@{}c@{}}Thick fog \\(Visibility range $<$ 40 m) \end{tabular} & \begin{tabular}[c]{@{}c@{}}Slight curve \\(R=310~m) \end{tabular} & \nonbox  & \nonbox & \donebox \\   \midrule
    6 & \begin{tabular}[c]{@{}c@{}}Thin fog \\(Visibility range $\geq$ 40 m) \end{tabular} & \begin{tabular}[c]{@{}c@{}}Sharp curve \\(R=170~m) \end{tabular} & \nonbox  & \nonbox & \donebox \\   \midrule
    7 & \begin{tabular}[c]{@{}c@{}}Thick fog \\(Visibility range $<$ 40 m) \end{tabular}  & \begin{tabular}[c]{@{}c@{}}Slight curve \\(R=320~m) \end{tabular} & \nonbox  & \nonbox & \donebox \\   \midrule
    8 & \begin{tabular}[c]{@{}c@{}}Thin fog \\(Visibility range $\geq$ 40 m) \end{tabular}  & \begin{tabular}[c]{@{}c@{}}Slight curve \\(R=330~m) \end{tabular} & \donebox & \nonbox &\nonbox\\   \midrule
    9 & \begin{tabular}[c]{@{}c@{}}Thin fog \\(Visibility range $\geq$ 40 m) \end{tabular}  & \begin{tabular}[c]{@{}c@{}}Sharp curve \\(R=160~m) \end{tabular} & \nonbox  & \nonbox & \donebox \\   \midrule
    10 & \begin{tabular}[c]{@{}c@{}}Clear \\(Visibility range $\geq$ 40 m) \end{tabular} & \begin{tabular}[c]{@{}c@{}}Sharp curve \\(R=150~m) \end{tabular} & \nonbox  & \nonbox & \donebox \\   \midrule
    11 & \begin{tabular}[c]{@{}c@{}}Thick fog \\(Visibility range $<$ 40 m) \end{tabular}  & Straight road  & \nonbox  & \nonbox & \donebox \\   \midrule
    12 & \begin{tabular}[c]{@{}c@{}}Clear \\(Visibility range $\geq$ 40 m) \end{tabular} & \begin{tabular}[c]{@{}c@{}}Slight curve \\(R=330~m) \end{tabular} & \donebox & \nonbox &\nonbox \\    \midrule
    13 & \begin{tabular}[c]{@{}c@{}}Thin fog \\(Visibility range $\geq$ 40 m) \end{tabular}  & Straight road & \donebox & \nonbox &\nonbox\\
    \bottomrule
  \end{tabular}
\end{table}

\subsubsection{Number of collisions in the Test Phase}

To evaluate the educational effect of three HMIs, the number of participants who collided in the test phase, \ie the RtI failure scenario, was counted for three groups, separately.
This was because that if drivers have difficulty establishing a correct mental model of the ADS, they are likely to over-trust and overly rely on the system~\cite{liu2019driving}, which increases the risk of delayed intervention during the test phase, potentially leading to collisions.

\subsubsection{Take-over time to ADS OFF in the Test Phase}

\begin{figure}[t]
  \centering
  \includegraphics[width=1\linewidth,clip,trim=0 15 0 0]{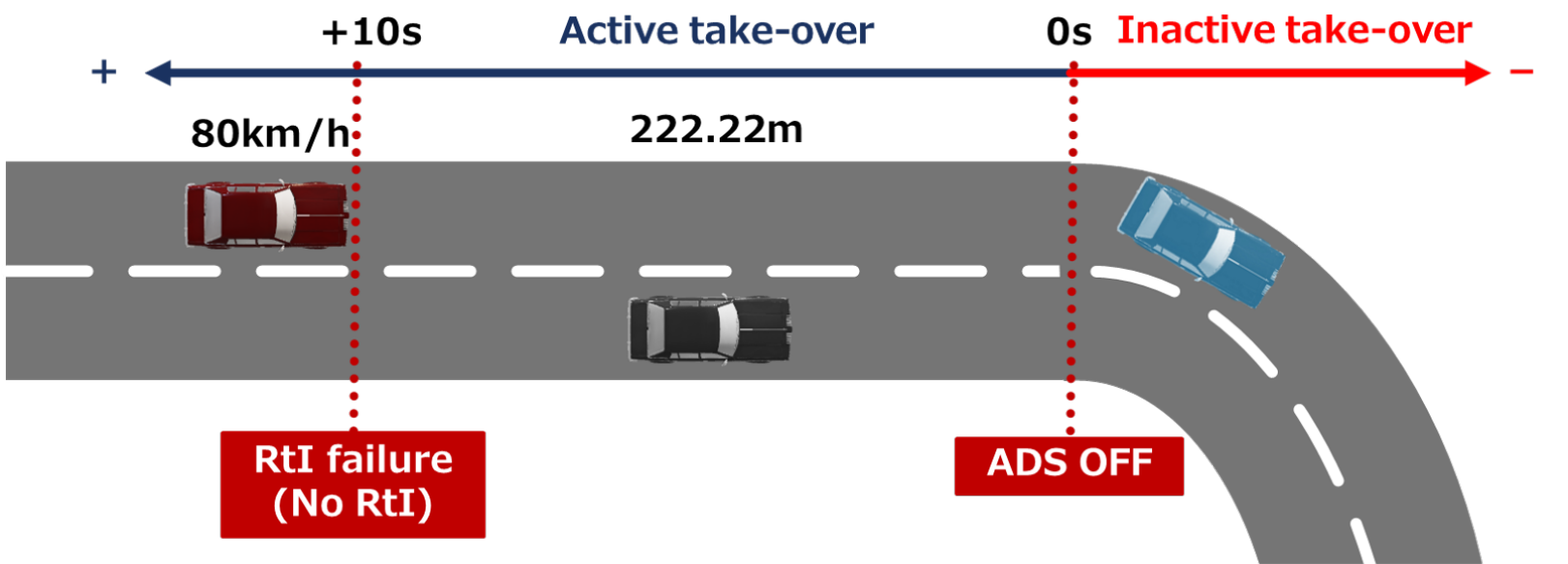}
  \caption{Illustration of an proactive take-over and a nonprotective take-over in the test phase, \ie the RtI failure scenario.}
  \label{fig:proactive_TO_image}
\end{figure}

For the same reason as counting the number of collisions, the take-over time in the test phase was also measured to evaluate the educational effect of the three HMIs.
Specifically, Fig.~\ref{fig:proactive_TO_image} shows an illustration of an proactive take-over and a nonprotective take-over in the test phase, \ie the RtI failure scenario.
The time points at which the take-over occurred by three groups of participants in the RtI failure scenario were measured.
The time point was set at which the timing of ADS deactivation (ADS OFF) as the 0-point (0~[s]).
The time before the 0-point are defined as positive values, while times after the 0-point are defined as negative values.
Furthermore, if the driver takes over before the 0-point, it is classified as an proactive take-over.
On the other hand, if the driver takes over after the 0-point, it is considered a nonprotective take-over, which significantly increases the risk of collision with the guardrail.

\section{Results}
\label{sec:Results}

All participants in the \textit{w/ trigger cue \& reason} group were confirmed that they received the RtI reason explanation when they proactively re-engaged the ADS after their takeover $35.3 \pm 8.24$~s in the RtI-issued scenarios of Learning Phases~I and II.
This ensured that the proposed method was correctly applied in the experiment.
For comparison, the \textit{w/o trigger cue} group and the \textit{w/ trigger cue} group re-engaged the ADS after their takeover $38.13 \pm 8.83$~s and $38.13 \pm 8.94$~s, respectively.

\subsection{Weighted Workload via NASA-TLX}
\begin{figure}[!h]
    \centering
    \includegraphics[width=1\linewidth]{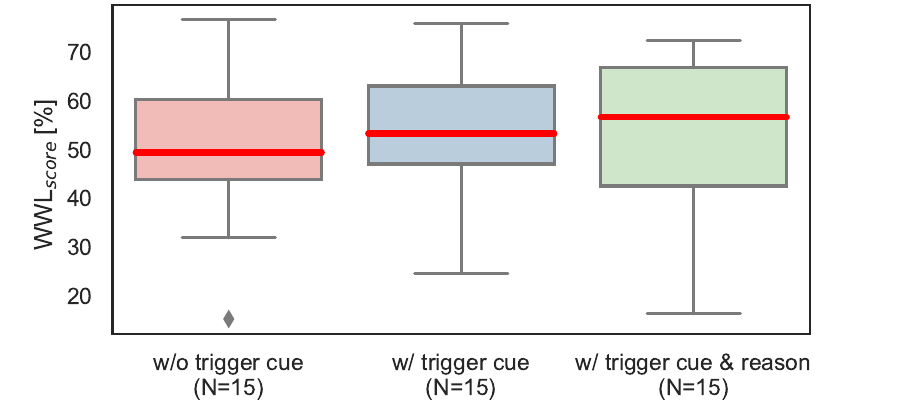}
    \caption{Workloads via NASA-TLX (WWL$_{scores}$)}
    \label{fig:NASA-TLX}
    %\vspace{-4mm}
\end{figure}

Fig.~\ref{fig:NASA-TLX} shows the WWL$_\text{score}$ of participants measured by the NASA-TLX.
The Shapiro–Wilk test showed non-normality ($W=0.872$, $p<0.001$), while the Levene test confirmed equal variance ($W=0.776$, $p=0.467$) among the three groups.
Accordingly, the Kruskal–Wallis H-test revealed no significant differences  ($H=1.702$, $p=0.427$) among them.

\subsection{Post-experiment Comprehension Test}

\begin{figure}[t]
  \centering
  \includegraphics[width=1\linewidth]{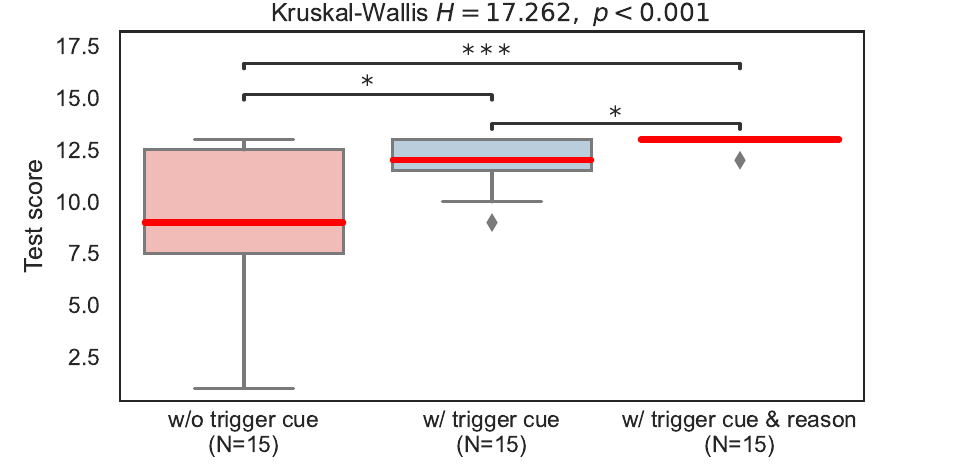}
  \caption{Post-experiment comprehension test results of three participant groups. $*: p<0.05$, $***: p<0.001$ via Mann-Whitney U test Bonferroni correction.}
  \label{fig:Test}
 
   \vspace{2mm}

\footnotesize
\setlength\tabcolsep{1.7pt}
\captionof{table}{Mann-Whitney U test results corrected by Bonferroni for Post-experiment comprehension test. $*: p<0.05$, $***: p<0.001$}
\label{tab:result_comprehension_post_hoc}
\centering
\begin{tabular}{@{}llrlr@{}}
\toprule
Group A & Group B & \multicolumn{1}{c}{\textit{U}} & \multicolumn{1}{c}{\textit{p}-adj.} & hedges' g \\ \midrule
w/ trigger cue           & w/ trigger cue \& reason & 58.0              & 0.017 *           & -1.012 \\
w/ trigger cue           & w/o trigger cue          & 169.5             & 0.047 *           & 1.055  \\
w/ trigger cue \& reason & w/o trigger cue          & 192.5             & 0.001 ***          & 1.534  \\ \bottomrule
\end{tabular}
%\vspace{-4mm}
\end{figure}

Fig.~\ref{fig:Test} presents the results of the post-experiment comprehension test.
The results indicate that the median test score for the group \textit{w/ trigger cue \& reason} and \textit{w/ trigger cue}, \textit{w/o trigger cue} groups were 13, 12 and 9 points, respectively. 
The Shapiro-Wilk test ($W=0.695$, $p<0.001$) and the Levene test ($W=12.746$, $p<0.001$) indicate that the results are non-normal and unequal variance.
Therefore, the result of the Kruskal-Wallis H-test ($H=17.262$, $p<0.001$) suggests significant differences among the three groups.
The results of post-hoc test via Mann–Whitney U test with Bonferroni correction are shown in Table~\ref{tab:result_comprehension_post_hoc}.

\subsection{Number of Collisions in the Test Phase}

\begin{figure}[h]
  \centering
  \includegraphics[width=1\linewidth]{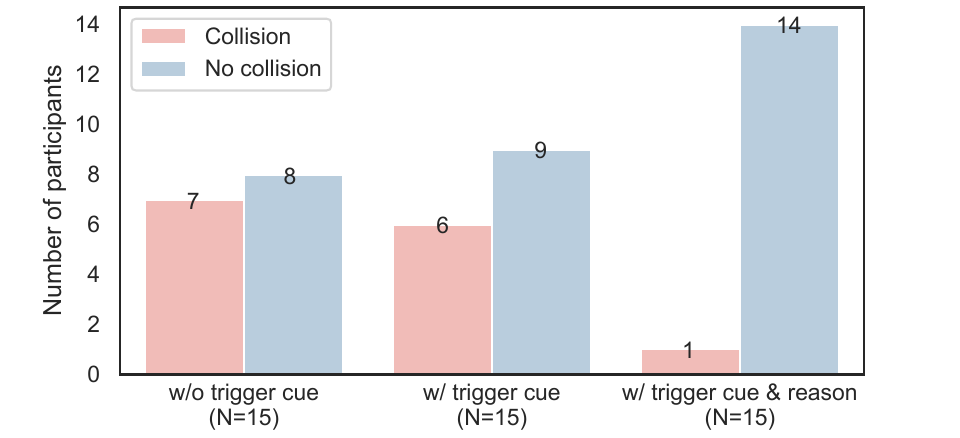}
 % \vspace{-4mm}
  \caption{Number of participants in three groups who had a collision in the test phase, \ie the RtI failure scenario.}
  \label{fig:collision}
  %\vspace{-4mm}
\end{figure}

Fig.~\ref{fig:collision} presents the number of collisions in the test phase, \ie RtI failure scenario.
%Specifically, in the \textit{w/ trigger cue \& reason} group, fourteen participants did not collide, but one had.
%In the \textit{w/ trigger cue} group, nine had no collisions, while six had.
%In the \textit{w/o trigger cue} group, eight had no collisions, while seven had.
A $2\times3$ Fisher's exact test suggested that there was a significant association among the number of collisions in three groups ($p=0.044$).

%\clearpage

\subsection{Take-over Time to ADS OFF in the Test Phase}

\begin{figure}[t]
 \centering
  \includegraphics[width=1\linewidth]{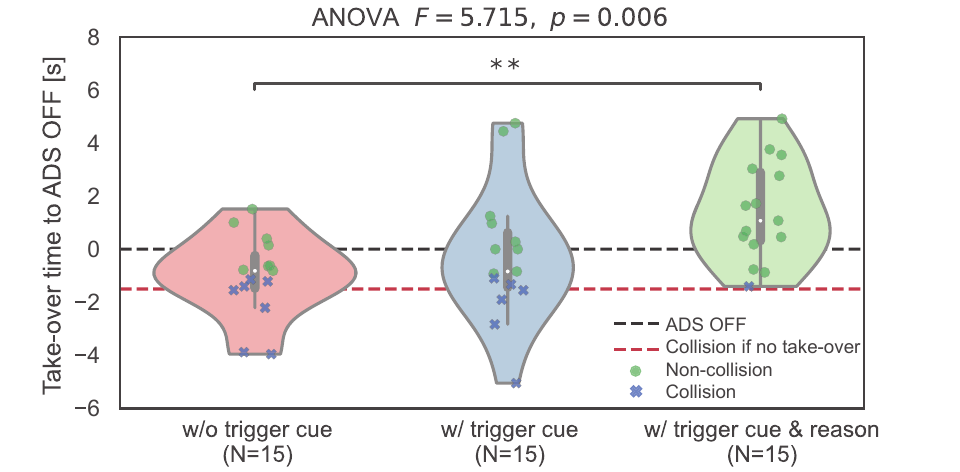}
  %\vspace{-2mm}
\caption{Take-over time to ADS OFF in the test phase (RtI failure scenario). Black dashed line: ADS OFF timing. Red dashed line: collision timing about 1.5 s after ADS OFF if no take-over. **: $p<0.01$ via Tukey-HSD test.}
  \label{fig:Take-over time to ADS OFF}
  
  \vspace{2mm}

\footnotesize
\setlength\tabcolsep{1.5pt}
\captionof{table}{Pairwise Comparison via Tukey-HSD test for the take-over time. $**: p<0.01$}
\label{tab:result_take_over_time_post_hoc}
\centering
\begin{tabular}{@{}llrlr@{}}
\toprule
Group A & Group B & \multicolumn{1}{c}{\textit{T}} & \multicolumn{1}{c}{\textit{p}-tukey} & hedges' g \\ \midrule
w/ trigger cue           & w/ trigger cue \& reason & -2.275            & 0.070        & -0.735\\
w/ trigger cue           & w/o trigger cue          & 1.028             & 0.564           & 0.353\\
w/ trigger cue \& reason & w/o trigger cue          & 3.303             & 0.005 **        & 1.381\\ \bottomrule
\end{tabular}
%\vspace{-2mm}
\end{figure}

Fig.~\ref{fig:Take-over time to ADS OFF} presents the results of the take-over time to ADS OFF in the RtI failure scenario. 
The red dashed line represents the collision occurring about 1.5~s after ADS OFF if the driver does not take over. 
Thus, take-over times later than this threshold indicate that the driver took over after the collision had already occurred.

The Shapiro-Wilk test ($W=0.966$, $p=0.204$) and the Levene test ($W=0.907$, $p=0.412$) indicated that the take-over time data followed a normal distribution and exhibited equal variance.
Therefore, an ANOVA was conducted, and its result ($F=5.715$, $p=0.006$) suggested significant differences in the take-over time among the three groups.
Furthermore, the results of the post hoc pairwise comparison for the take-over time among the three groups using the Tukey-HSD test are shown in Table~\ref{tab:result_take_over_time_post_hoc}. 

\begin{figure}[h]
  \centering
  \includegraphics[width=1\linewidth]{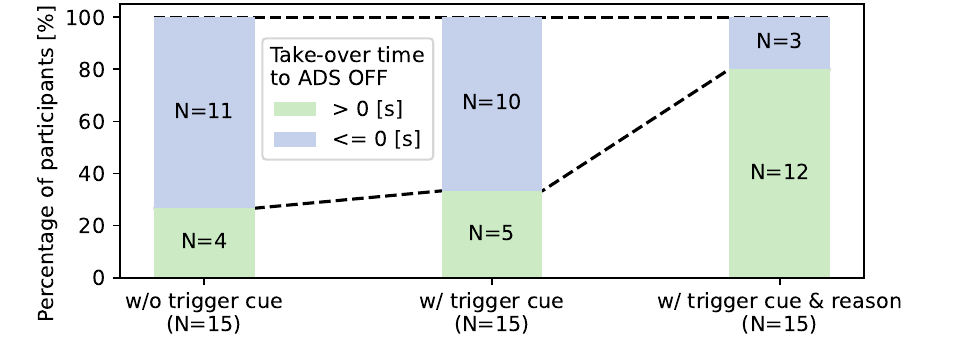}
  %\vspace{-2mm}
  \caption{Number of participants who proactively take-over in the test phase, \ie the RtI failure scenario.}
  \label{fig:Number of participants whose take-over time}
  %\vspace{-2mm}
\end{figure}

Fig.~\ref{fig:Number of participants whose take-over time} presents the results showing the number of participants who proactively took over before ADS OFF in the RtI failure scenario, as well as those who did not.
These results suggest that most of the participants in the \textit{w/ trigger cue \& reason} group took over control more proactively before the ADS OFF, while the majority of participants in the other groups did not exhibit the same behavior.

\subsection{Correlation Between the Post-Experiment Comprehension Test Scores and the Take-Over Time to the ADS OFF}

\begin{figure}[ht]
 \centering
  \includegraphics[width=0.85\linewidth]{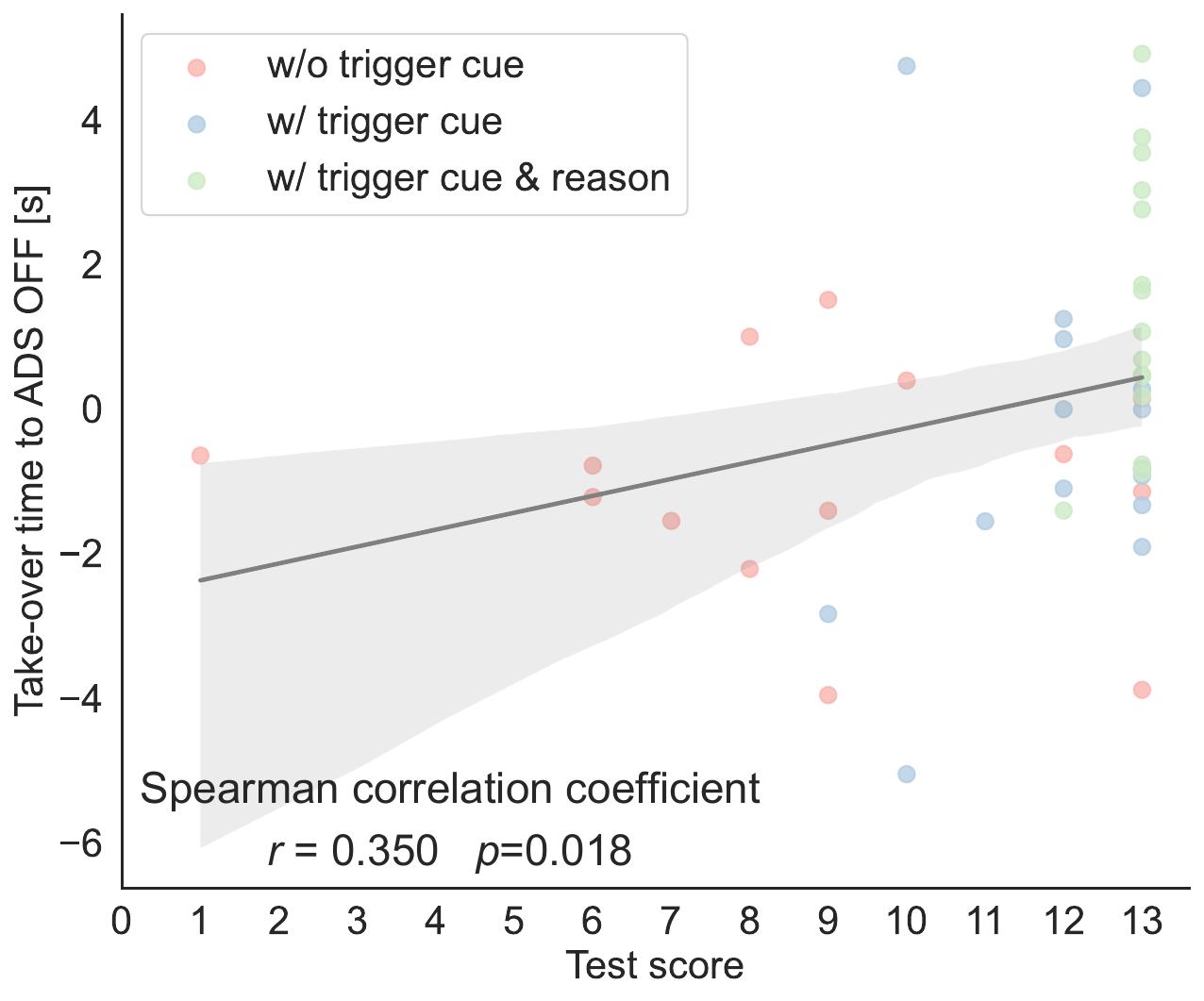}
    \vspace{-2mm}
  \caption{Correlation between post-experiment comprehension test score and take-over time to ADS OFF in the RtI failure scenario  (h).}
  \label{fig:Correlation}
  \vspace{-4mm}
\end{figure}

To examine the association between post-experiment comprehension test scores and take-over time relative to ADS OFF, the results of the Spearman rank-order correlation analysis are presented in Fig.~\ref{fig:Correlation} and Table~\ref{tab:result_Correlation}.

\begin{table}[!h]
\footnotesize
\setlength\tabcolsep{4pt}
\captionof{table}{Spearman rank-order correlation between post-experiment comprehension test score and take-over time to ADS OFF in the RtI failure scenario (h). $*: p<0.05$}
\label{tab:result_Correlation}
\centering
\begin{tabular}{crrrrrrl@{}}
\toprule
&  \multicolumn{2}{c}{\begin{tabular}[c]{@{}c@{}}\footnotesize{Post-experiment} \\ \footnotesize{comprehension test}\end{tabular}}
& \multicolumn{2}{c}{\begin{tabular}[c]{@{}c@{}}\footnotesize{Take-over time} \\ \footnotesize{to ADS OFF}\end{tabular}} 
&\multicolumn{3}{c}{\begin{tabular}[c]{@{}c@{}}\footnotesize{Spearman rank-order} \\ \footnotesize{correlation}\end{tabular}}  \\ 
\cmidrule(l){2-3} \cmidrule(l){4-5} \cmidrule(l){6-8} 
 N & \multicolumn{1}{r}{Ave.} & \multicolumn{1}{r}{Std.} & \multicolumn{1}{r}{Ave.} & \multicolumn{1}{r}{Std.} & \textit{r} & CI95\% & \textit{p}-values\\ 
\midrule
 45 & 11.333           & 2.629            & 0.049    & 2.216    & 0.350 & [0.06, 0.58] & 0.018 * \\ \bottomrule
\end{tabular}
%\vspace{-2mm}
\end{table}

\section{Discussion}
\label{sec:Discussion}

\subsection{Weighted Workload via NASA-TLX}
The result of the weighted workload (WWL$_{scores}$) evaluated by NASA-TLX  indicates no differences among the three groups as shown in Fig.~\ref{fig:NASA-TLX}.
Thus, it suggests that the proposed HMI, which provides additional information during the RtI as observed in the \textit{w/ trigger cue \& reason} group, did not adversely impact drivers' workload in answering RQ~1.

\subsection{Post-experiment Comprehension Test}

For RQ~2, the comprehension test presented in Fig.~\ref{fig:Test} and Table~\ref{table:Post-experiment comprehension test} indicates that the test scores of the group that had the trigger cue \& reason on RtI were significantly higher than those of the \textit{w/o trigger cue} group and the \textit{w/ trigger cue} group.
The results suggest that drivers will correctly understand the system limitations of ADS through repeated use of the proposed HMI.
Furthermore, the high scores on the post-experiment comprehension test for the \textit{w/ trigger cue \& reason} group also suggest that the proposed RtI HMI had a sustained educational effect compared to the pre-educational method (i.e., the pre-instructions received by the \textit{w/o trigger cue} group).
This finding aligns with the conclusion in~\cite{merriman2023does} which showed that the driver's training affects their mental models, \ie their comprehension of system limitations.

\subsection{Number of Collisions and Take-over Timing in Test Phase}

Addressing RQ3, we interpret the results to indicate that repeated use of the \textit{RtI HMI w/ trigger cue \& reason} fosters proactive driver intervention and reduces accidents.

In the test Phase, \ie the RtI failure scenario, relative to the \textit{w/o trigger cue} condition, the \textit{w/ trigger cue} condition yielded a small decrease in collisions (Fig.~\ref{fig:collision}), aligning with prior studies suggesting that environment-linked cues can modestly strengthen situational awareness~\cite{wright2018effective}.
However, no significant difference in take-over timing was observed between these two groups (see Table~\ref{tab:result_take_over_time_post_hoc}).

Moreover, as shown in Fig.~\ref{fig:collision}, the \textit{w/ trigger cue \& reason} group exhibited a substantially lower number of collisions than the other two groups in the RtI failure scenario.
One possible explanation for this reduction is that more participants in this group proactively took over control (Fig.~\ref{fig:Number of participants whose take-over time}), and that they tended to take over earlier than those in the \textit{w/ trigger cue} group, and significantly earlier than those in the \textit{w/o trigger cue} group (Fig.~\ref{fig:Take-over time to ADS OFF}, Table~\ref{tab:result_take_over_time_post_hoc}).
These results suggest that presenting trigger cues with reason explanations was associated with earlier take-over behavior in boundary situations, which may contribute to avoiding potential collisions.
These findings are consistent with related studies~\cite{zhou2021effects, zhou2021influence}, which demonstrated that a clearer understanding of system limitations facilitates earlier and safer take-over behavior.

In summary, repeated use of the \textit{RtI HMI w/ trigger cue \& reason} was associated with learning-related improvements in drivers' understanding of ADS system limitations, earlier takeovers by drivers prior to ADS deactivation, and lower accident occurrence, addressing RQ~3.

\subsection{Correlation between the Post-experiment Comprehension Test and the Take-over Timing}

To further explore the cognitive factors potentially underlying the observed differences in take-over timing across the three groups, we examined the relationship between take-over timing and drivers' comprehension of ADS limitations.
Fig.~\ref{fig:Correlation} and Table~\ref{tab:result_Correlation} show a significant positive correlation ($r=0.35$, $p<0.05$) between post-experiment comprehension test scores and take-over time to ADS OFF.
This correlation indicates that a higher level of understanding of ADS limitations was associated with earlier proactive take-over behavior in the RtI failure scenario, that is, before ADS deactivation, thereby addressing RQ~4.
These results are consistent with prior findings~\cite{zhou2021does,zhou2021effects}, suggesting that drivers' knowledge of ADS functions may influence take-over behavior during system failures.
Note that this correlational finding does not support causal interpretation, and the presence of unmeasured confounding factors, \eg driving skill and trust in the ADS, cannot be excluded.
The mechanisms linking drivers' comprehension of ADS limitations to take-over behavior remain to be clarified.

\subsection{Limitations and Future Works}

Although the experiment involved 45 participants divided into three groups, the limited sample size and the predominance of young Japanese participants may restrict the generalizability of the findings to drivers with different cultural backgrounds, driving experiences, or age groups.
Future work will recruit more diverse participants to enhance the generalizability of the results.

The sequence of scenarios was intentionally fixed to minimize confounding effects among groups and enhance comparability. 
However, the formation of participants' mental models of the ADS may vary with scenario order, and this fixed design may have introduced potential order effects.
Future work will adopt counterbalanced scenario orders.

As the SAE Level 3 ADS allows but does not require the driver to perform NDRTs, participants in this experiment were not instructed to engage in NDRTs. 
However, this setting may limit the generalization of the findings when drivers are occupied with NDRTs before the RtI.
Future work will evaluate the proposed method under realistic NDRT conditions.

This study was designed with the RtI triggered 10 s before the ADS exceeded its system limitations, as drivers typically need about 10 s to safely take over control~\citep{merat2014transition}.
However, whether the educational effects of the proposed RtI HMI remain effective in sudden or emergency situations is still unclear.
Future work will evaluate these effects under different take-over time budgets.

At last, differences between the experimental simulator and real-world conditions, such as the ODD of level 3 AVs and driving environments, cannot be ignored.

\section{Conclusion}
\label{sec:Conclusion}
This study proposed a HMI providing the RtI trigger cues \& reason, to help drivers correctly understand the system limitation of an ADS in scenarios.
The results of a between-group experiment using a driving simulator showed that the use of trigger cues and reason explanations was associated with better driver comprehension of system limitations and earlier takeovers by drivers during RtI failure scenarios, as well as lower collision occurrence.
These findings suggested that trigger cues and reason explanations are valuable for helping drivers to comprehend system limitations of the ADS.

\section*{Acknowledgments}
This work was supported by JSPS KAKENHI Grant Numbers 20K19846 and 22H00246, Japan.

\section*{CRediT Author Statement}
\noindent
\textbf{Ryuji~Matsuo}: Methodology, Software, Investigation, Formal Analysis, Visualization, Writing - Original Draft.\\
\textbf{Hailong~Liu}: Conceptualization, Methodology, Formal Analysis, Visualization, Supervision, Project administration, Funding acquisition, Writing - Original Draft \& review \& editing.\\
\textbf{Toshihiro~Hiraoka}: Methodology, Writing - review \& editing.\\
\textbf{Takahiro Wada}:  Methodology, Writing - review \& editing.

\footnotesize
\bibliographystyle{IEEEtranN} 
\bibliography{sample.bib}

% Generated by IEEEtranN.bst, version: 1.14 (2015/08/26)
\begin{thebibliography}{42}
\providecommand{\natexlab}[1]{#1}
\providecommand{\url}[1]{#1}
\csname url@samestyle\endcsname
\providecommand{\newblock}{\relax}
\providecommand{\bibinfo}[2]{#2}
\providecommand{\BIBentrySTDinterwordspacing}{\spaceskip=0pt\relax}
\providecommand{\BIBentryALTinterwordstretchfactor}{4}
\providecommand{\BIBentryALTinterwordspacing}{\spaceskip=\fontdimen2\font plus
\BIBentryALTinterwordstretchfactor\fontdimen3\font minus \fontdimen4\font\relax}
\providecommand{\BIBforeignlanguage}[2]{{%
\expandafter\ifx\csname l@#1\endcsname\relax
\typeout{** WARNING: IEEEtranN.bst: No hyphenation pattern has been}%
\typeout{** loaded for the language `#1'. Using the pattern for}%
\typeout{** the default language instead.}%
\else
\language=\csname l@#1\endcsname
\fi
#2}}
\providecommand{\BIBdecl}{\relax}
\BIBdecl

\bibitem[Yurtsever et~al.(2020)Yurtsever, Lambert, Carballo, and Takeda]{9046805}
E.~Yurtsever, J.~Lambert, A.~Carballo, and K.~Takeda, ``A survey of autonomous driving: Common practices and emerging technologies,'' \emph{IEEE Access}, vol.~8, pp. 58\,443--58\,469, 2020.

\bibitem[Li et~al.(2021)Li, Cheng, Zeng, Liu, and Sester]{li2021autonomous}
Y.~Li, H.~Cheng, Z.~Zeng, H.~Liu, and M.~Sester, ``Autonomous vehicles drive into shared spaces: ehmi design concept focusing on vulnerable road users,'' in \emph{2021 IEEE International Intelligent Transportation Systems Conference}.\hskip 1em plus 0.5em minus 0.4em\relax IEEE, 2021, pp. 1729--1736.

\bibitem[{SAE International}(2018)]{sae2018taxonomy}
{SAE International}, ``Taxonomy and definitions for terms related to driving automation systems for on-road motor vehicles,'' \emph{SAE international}, vol. 4970, no. 724, pp. 1--5, 2018.

\bibitem[Merat et~al.(2014)Merat, Jamson, Lai, Daly, and Carsten]{merat2014transition}
N.~Merat, A.~H. Jamson, F.~C. Lai, M.~Daly, and O.~M. Carsten, ``Transition to manual: Driver behaviour when resuming control from a highly automated vehicle,'' \emph{Transportation research part F: traffic psychology and behaviour}, vol.~27, pp. 274--282, 2014.

\bibitem[Strand et~al.(2014)Strand, Nilsson, Karlsson, and Nilsson]{strand2014semi}
N.~Strand, J.~Nilsson, I.~M. Karlsson, and L.~Nilsson, ``Semi-automated versus highly automated driving in critical situations caused by automation failures,'' \emph{Transportation research part F: traffic psychology and behaviour}, vol.~27, pp. 218--228, 2014.

\bibitem[Saito et~al.(2018)Saito, Wada, and Sonoda]{saito2018control}
T.~Saito, T.~Wada, and K.~Sonoda, ``Control authority transfer method for automated-to-manual driving via a shared authority mode,'' \emph{IEEE Transactions on Intelligent Vehicles}, vol.~3, no.~2, pp. 198--207, 2018.

\bibitem[Kondo et~al.(2019{\natexlab{a}})Kondo, Wada, and Sonoda]{kondo2019use}
R.~Kondo, T.~Wada, and K.~Sonoda, ``Use of haptic shared control in highly automated driving systems,'' \emph{IFAC-PapersOnLine}, vol.~52, no.~19, pp. 43--48, 2019.

\bibitem[Okada et~al.(2020)Okada, Sonoda, and Wada]{okada2020transferring}
K.~Okada, K.~Sonoda, and T.~Wada, ``Transferring from automated to manual driving when traversing a curve via haptic shared control,'' \emph{IEEE Transactions on Intelligent Vehicles}, vol.~6, no.~2, pp. 266--275, 2020.

\bibitem[Kondo et~al.(2019{\natexlab{b}})Kondo, Sonoda, and Wada]{kondo2019shared}
R.~Kondo, K.~Sonoda, and T.~Wada, ``Shared authority mode in uncertain situation,'' in \emph{2019 IEEE International Conference on Systems, Man and Cybernetics}.\hskip 1em plus 0.5em minus 0.4em\relax IEEE, 2019, pp. 3110--3115.

\bibitem[Liu et~al.(2019)Liu, Hiraoka, and Tanaka]{liu2019overtrust}
H.~Liu, T.~Hiraoka, and S.~Tanaka, ``Explicit behaviors affected by driver's trust in a driving automation system,'' in \emph{The 5th International Symposium on Future Active Safety Technology toward Zero Accidents}, 2019, pp. 1--6.

\bibitem[Liu and Hiraoka(2019)]{liu2019driving}
H.~Liu and T.~Hiraoka, ``Driving behavior model considering driver's over-trust in driving automation system,'' in \emph{Proceedings of the 11th International Conference on Automotive User Interfaces and Interactive Vehicular Applications: Adjunct Proceedings}, 2019, pp. 115--119.

\bibitem[Zhou et~al.(2021{\natexlab{a}})Zhou, Itoh, and Kitazaki]{zhou2021does}
H.~Zhou, M.~Itoh, and S.~Kitazaki, ``How does explanation-based knowledge influence driver take-over in conditional driving automation?'' \emph{IEEE Transactions on Human-Machine Systems}, vol.~51, no.~3, pp. 188--197, 2021.

\bibitem[Staggers and Norcio(1993)]{staggers1993mental}
N.~Staggers and A.~F. Norcio, ``Mental models: concepts for human-computer interaction research,'' \emph{International Journal of Man-machine studies}, vol.~38, no.~4, pp. 587--605, 1993.

\bibitem[Liu et~al.(2021)Liu, Hirayama, and Watanabe]{liu2021importance}
H.~Liu, T.~Hirayama, and M.~Watanabe, ``Importance of instruction for pedestrian-automated driving vehicle interaction with an external human machine interface: Effects on pedestrians' situation awareness, trust, perceived risks and decision making,'' in \emph{2021 IEEE Intelligent Vehicles Symposium}.\hskip 1em plus 0.5em minus 0.4em\relax IEEE, 2021, pp. 748--754.

\bibitem[Liu and Hirayama(2025)]{liu2025}
H.~Liu and T.~Hirayama, ``Pre-instruction for pedestrians interacting autonomous vehicles with {eHMI}: Effects on their psychology and walking behavior,'' \emph{IEEE Transactions on Intelligent Transportation Systems}, vol.~26, no.~8, pp. 11\,313--11\,324, 2025.

\bibitem[Schwalk et~al.(2015)Schwalk, Kalogerakis, and Maier]{schwalk2015driver}
M.~Schwalk, N.~Kalogerakis, and T.~Maier, ``Driver support by a vibrotactile seat matrix--recognition, adequacy and workload of tactile patterns in take-over scenarios during automated driving,'' \emph{Procedia Manufacturing}, vol.~3, pp. 2466--2473, 2015.

\bibitem[Borojeni et~al.(2017)Borojeni, Wallbaum, Heuten, and Boll]{borojeni2017comparing}
S.~S. Borojeni, T.~Wallbaum, W.~Heuten, and S.~Boll, ``Comparing shape-changing and vibro-tactile steering wheels for take-over requests in highly automated driving,'' in \emph{Proceedings of the 9th international conference on automotive user interfaces and interactive vehicular applications}, 2017, pp. 221--225.

\bibitem[Morales-Alvarez et~al.(2022)Morales-Alvarez, Certad, Tadjine, and Olaverri-Monreal]{morales2022automated}
W.~Morales-Alvarez, N.~Certad, H.~H. Tadjine, and C.~Olaverri-Monreal, ``Automated driving systems: Impact of haptic guidance on driving performance after a take over request,'' in \emph{2022 IEEE Intelligent Vehicles Symposium}.\hskip 1em plus 0.5em minus 0.4em\relax IEEE, 2022, pp. 1817--1823.

\bibitem[Politis et~al.(2015)Politis, Brewster, and Pollick]{Politis2015}
I.~Politis, S.~Brewster, and F.~Pollick, ``To beep or not to beep? comparing abstract versus language-based multimodal driver displays,'' \emph{\textit{the 33rd Annual ACM Conference on Human Factors in Computing Systems}}, p. 3971^^e2^^80^^933980, 2015.

\bibitem[Forster et~al.(2017)Forster, Naujoks, Neukum, and Huestegge]{forster2017driver}
Y.~Forster, F.~Naujoks, A.~Neukum, and L.~Huestegge, ``Driver compliance to take-over requests with different auditory outputs in conditional automation,'' \emph{Accident Analysis \& Prevention}, vol. 109, pp. 18--28, 2017.

\bibitem[El~Jouhri et~al.(2023)El~Jouhri, El~Sharkawy, Paksoy, Youssif, He, Kim, and Happee]{el2023influence}
A.~El~Jouhri, A.~El~Sharkawy, H.~Paksoy, O.~Youssif, X.~He, S.~Kim, and R.~Happee, ``The influence of a color themed hmi on trust and take-over performance in automated vehicles,'' \emph{Frontiers in psychology}, vol.~14, p. 1128285, 2023.

\bibitem[Gon{\c{c}}alves et~al.(2023)Gon{\c{c}}alves, Louw, Lee, Madigan, Kuo, Lenn{\'e}, and Merat]{gonccalves2023users}
R.~C. Gon{\c{c}}alves, T.~Louw, Y.~M. Lee, R.~Madigan, J.~Kuo, M.~Lenn{\'e}, and N.~Merat, ``Is users’ trust during automated driving different when using an ambient light hmi, compared to an auditory hmi?'' \emph{Information}, vol.~14, no.~5, p. 260, 2023.

\bibitem[Ou et~al.(2021)Ou, Huang, and Fang]{ou2021effects}
Y.-K. Ou, W.-X. Huang, and C.-W. Fang, ``Effects of different takeover request interfaces on takeover behavior and performance during conditionally automated driving,'' \emph{Accident Analysis \& Prevention}, vol. 162, p. 106425, 2021.

\bibitem[Borojeni et~al.(2016)Borojeni, Chuang, Heuten, and Boll]{Borojeni2016}
S.~S. Borojeni, L.~Chuang, W.~Heuten, and S.~Boll, ``Assisting drivers with ambient take-over requests in highly automated driving,'' \emph{\textit{AutomotiveUI'16}}, 2016.

\bibitem[Wright et~al.(2018)Wright, Agrawal, Samuel, Wang, Zilberstein, and Fisher]{wright2018effective}
T.~J. Wright, R.~Agrawal, S.~Samuel, Y.~Wang, S.~Zilberstein, and D.~L. Fisher, ``Effective cues for accelerating young drivers’ time to transfer control following a period of conditional automation,'' \emph{Accident Analysis \& Prevention}, vol. 116, pp. 14--20, 2018.

\bibitem[Endsley(1995)]{endsley1995toward}
M.~R. Endsley, ``Toward a theory of situation awareness in dynamic systems,'' \emph{Human factors}, vol.~37, no.~1, pp. 32--64, 1995.

\bibitem[Jones et~al.(2011)Jones, Ross, Lynam, Perez, and Leitch]{jones2011mental}
N.~A. Jones, H.~Ross, T.~Lynam, P.~Perez, and A.~Leitch, ``Mental models: an interdisciplinary synthesis of theory and methods,'' \emph{Ecology and society}, vol.~16, no.~1, 2011.

\bibitem[Boelhouwer et~al.(2019)Boelhouwer, van~den Beukel, van~der Voort, and Martens]{boelhouwer2019should}
A.~Boelhouwer, A.~P. van~den Beukel, M.~C. van~der Voort, and M.~H. Martens, ``Should i take over? does system knowledge help drivers in making take-over decisions while driving a partially automated car?'' \emph{Transportation research part F: traffic psychology and behaviour}, vol.~60, pp. 669--684, 2019.

\bibitem[Ebnali et~al.(2019)Ebnali, Hulme, Ebnali-Heidari, and Mazloumi]{EBNALI2019184}
M.~Ebnali, K.~Hulme, A.~Ebnali-Heidari, and A.~Mazloumi, ``How does training effect users’ attitudes and skills needed for highly automated driving?'' \emph{Transportation Research Part F: Traffic Psychology and Behaviour}, vol.~66, pp. 184--195, 2019.

\bibitem[K{\"o}rber et~al.(2018)K{\"o}rber, Prasch, and Bengler]{korber2018have}
M.~K{\"o}rber, L.~Prasch, and K.~Bengler, ``Why do i have to drive now? post hoc explanations of takeover requests,'' \emph{Human factors}, vol.~60, no.~3, pp. 305--323, 2018.

\bibitem[Matsuo et~al.(2024)Matsuo, Liu, Hiraoka, and Wada]{Matsuo2024ICHMS}
R.~Matsuo, H.~Liu, T.~Hiraoka, and T.~Wada, ``Enhancing the driver’s comprehension of ads’s system limitations: An hmi providing request-to-intervene trigger and reason explanation,'' in \emph{2024 IEEE 4th International Conference on Human-Machine Systems}, 2024, pp. 1--7.

\bibitem[Petermeijer et~al.(2017)Petermeijer, Doubek, and De~Winter]{petermeijer2017driver}
S.~Petermeijer, F.~Doubek, and J.~De~Winter, ``Driver response times to auditory, visual, and tactile take-over requests: A simulator study with 101 participants,'' in \emph{2017 IEEE International Conference on Systems, Man, and Cybernetics}.\hskip 1em plus 0.5em minus 0.4em\relax IEEE, 2017, pp. 1505--1510.

\bibitem[{MLIT, Japan}(1970 (update: 2020))]{JP_RSO}
{MLIT, Japan}, ``Road structure ordinance~(\begin{CJK}{UTF8}{ipxm}道路構造令\end{CJK}),'' \url{https://laws.e-gov.go.jp/law/345CO0000000320}, 1970 (update: 2020).

\bibitem[Chaabani et~al.(2018)Chaabani, Werghi, Kamoun, Taha, Outay, and Yasar]{chaabani2018estimating}
H.~Chaabani, N.~Werghi, F.~Kamoun, B.~Taha, F.~Outay, and A.-U.-H. Yasar, ``Estimating meteorological visibility range under foggy weather conditions: A deep learning approach,'' \emph{Procedia Computer Science}, vol. 141, pp. 478--483, 2018.

\bibitem[Pereira et~al.(2024)Pereira, Cruz, Fernandes, Pinto, and Cardoso]{pereira2024weather}
C.~Pereira, R.~P. Cruz, J.~N. Fernandes, J.~R. Pinto, and J.~S. Cardoso, ``Weather and meteorological optical range classification for autonomous driving,'' \emph{IEEE Transactions on Intelligent Vehicles}, 2024.

\bibitem[Mori et~al.(2007)Mori, Takahashi, Ide, Murase, Miyahara, and Tamatsu]{mori2007fog}
K.~Mori, T.~Takahashi, I.~Ide, H.~Murase, T.~Miyahara, and Y.~Tamatsu, ``Fog density recognition by in-vehicle camera and millimeter wave radar,'' \emph{Int. J. Control}, vol.~3, no.~5, pp. 1173--1182, 2007.

\bibitem[Cao et~al.(2023)Cao, Wang, and Li]{cao2023fog}
R.~Cao, X.~Wang, and H.~Li, ``Fog density evaluation by combining image grayscale entropy and directional entropy,'' \emph{Atmosphere}, vol.~14, no.~7, p. 1125, 2023.

\bibitem[Hart and Staveland(1988)]{hart1988development}
S.~G. Hart and L.~E. Staveland, ``Development of nasa-tlx (task load index): Results of empirical and theoretical research,'' in \emph{Advances in psychology}.\hskip 1em plus 0.5em minus 0.4em\relax Elsevier, 1988, vol.~52, pp. 139--183.

\bibitem[Haga and Mizukami(1996)]{haga1996japanese}
S.~Haga and N.~Mizukami, ``Japanese version of nasa task load index sensitivity of its workload score to difficulty of three different laboratory tasks,'' \emph{The Japanese journal of ergonomics}, vol.~32, no.~2, pp. 71--79, 1996.

\bibitem[Merriman et~al.(2023)Merriman, Revell, and Plant]{merriman2023does}
S.~E. Merriman, K.~M. Revell, and K.~L. Plant, ``What does an automated vehicle class as a hazard? using online video-based training to improve drivers’ trust and mental models for activating an automated vehicle,'' \emph{Transportation research part F: traffic psychology and behaviour}, vol.~98, pp. 1--17, 2023.

\bibitem[Zhou et~al.(2021{\natexlab{b}})Zhou, Kamijo, Itoh, and Kitazaki]{zhou2021effects}
H.~Zhou, K.~Kamijo, M.~Itoh, and S.~Kitazaki, ``Effects of explanation-based knowledge regarding system functions and driver’s roles on driver takeover during conditionally automated driving: A test track study,'' \emph{Transportation research part F: traffic psychology and behaviour}, vol.~77, pp. 1--9, 2021.

\bibitem[Zhou et~al.(2021{\natexlab{c}})Zhou, Itoh, and Kitazaki]{zhou2021influence}
H.~Zhou, M.~Itoh, and S.~Kitazaki, ``Influence of prior general knowledge on older adults’ takeover performance and attitude toward using conditionally automated driving systems,'' in \emph{Proceedings of the Human Factors and Ergonomics Society Annual Meeting}, vol.~65, no.~1.\hskip 1em plus 0.5em minus 0.4em\relax SAGE Publications Sage CA: Los Angeles, CA, 2021, pp. 1327--1331.

\end{thebibliography}

\vspace{-11mm}
\begin{IEEEbiography}
[{\includegraphics[width=1in,height=1.25in,clip,keepaspectratio]{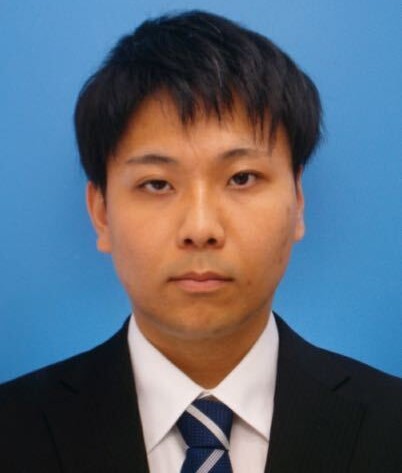}}]
{Ryuji~Matsuo}
 received his M.Eng. degree from the Graduate School of Information Science and Engineering, Graduate School of Science and Technology, NAIST, Japan, in 2024.
His research focused on human factors in the intelligent transportation systems.
\end{IEEEbiography}
\vspace{-9mm}

\begin{IEEEbiography}
[{\includegraphics[width=1in,height=1.25in,clip,keepaspectratio]{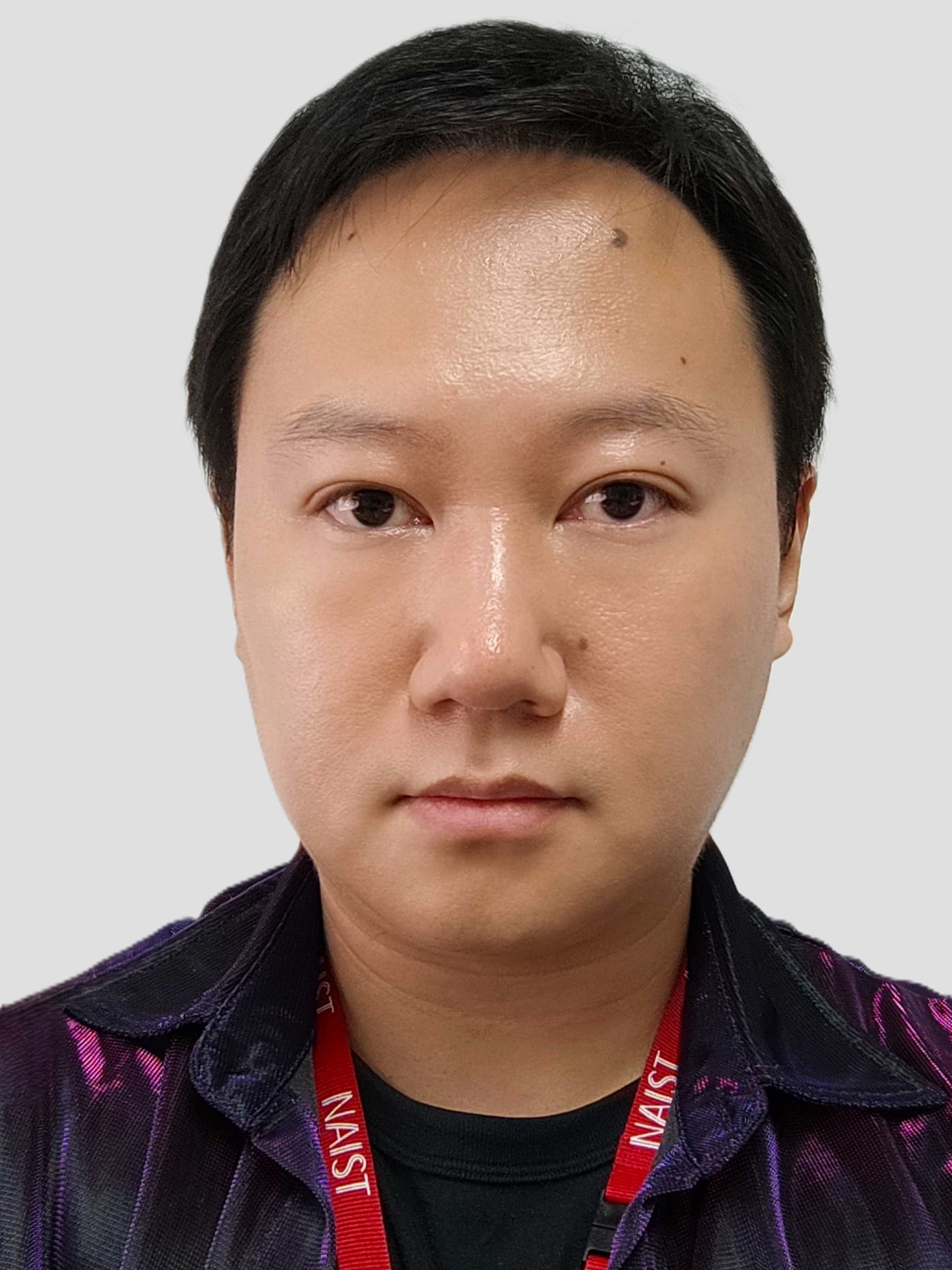}}]
{Hailong~Liu} (S'15--M'19--SM'25) received his B.Eng., M.Eng. and Ph.D. degrees in Engineering from Ritsumeikan University, Japan, in 2013, 2015 and 2018, respectively. He was a JSPS Research Fellow for Young Scientists (DC2) (2016--2018), a researcher at Nagoya University (2018--2021).
In Nov. 2021, he joined Nara Institute of Science and Technology (NAIST), Japan, as an Assistant Professor and was promoted to Associate Professor in Feb. 2024.
From Oct. 2025, he becomes a cancer warrior.
His research focuses on human factors and machine learning in intelligent transportation systems. He is a Senior Member of IEEE and holds memberships in IEEE ITSS, RAS, SMC. He also serves on the Human Factors in ITS Committee of IEEE ITSS. In addition, he is a member of JSAE, JSAI, and SICE.
\end{IEEEbiography}
\vspace{-9mm}

\begin{IEEEbiography}
[{\includegraphics[width=1in,height=1.25in,clip,keepaspectratio,trim=20 0 30 20]{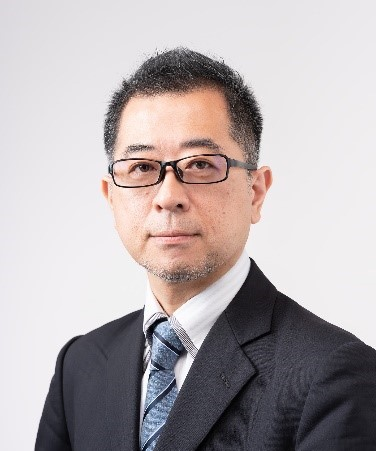}}]
{Toshihiro~Hiraoka} (M'14) received B.E. and M.E. degrees in Precision Engineering in 1994 and 1996, and a Ph.D. in Informatics in 2005, all from Kyoto University, Japan.
He worked at Matsushita Electric Industrial Co., Ltd. (1996--1998), Kyoto University as an Assistant Professor (1998--2017), Nagoya University as a Designated Associate Professor (2017--2019), and the University of Tokyo as a project professor (2019--2022). Since 2022, he has been a Senior Chief Researcher at the Japan Automobile Research Institute. His research interests include human-machine systems, advanced driver-assistance systems, and automated driving systems. He is a member of SICE, HIS, JSAE, JES, IATSS, and IEEE (ITSS).

 \end{IEEEbiography}
\vspace{-9mm}

\begin{IEEEbiography} [{\includegraphics[width=1in,height=1.25in,clip,keepaspectratio]{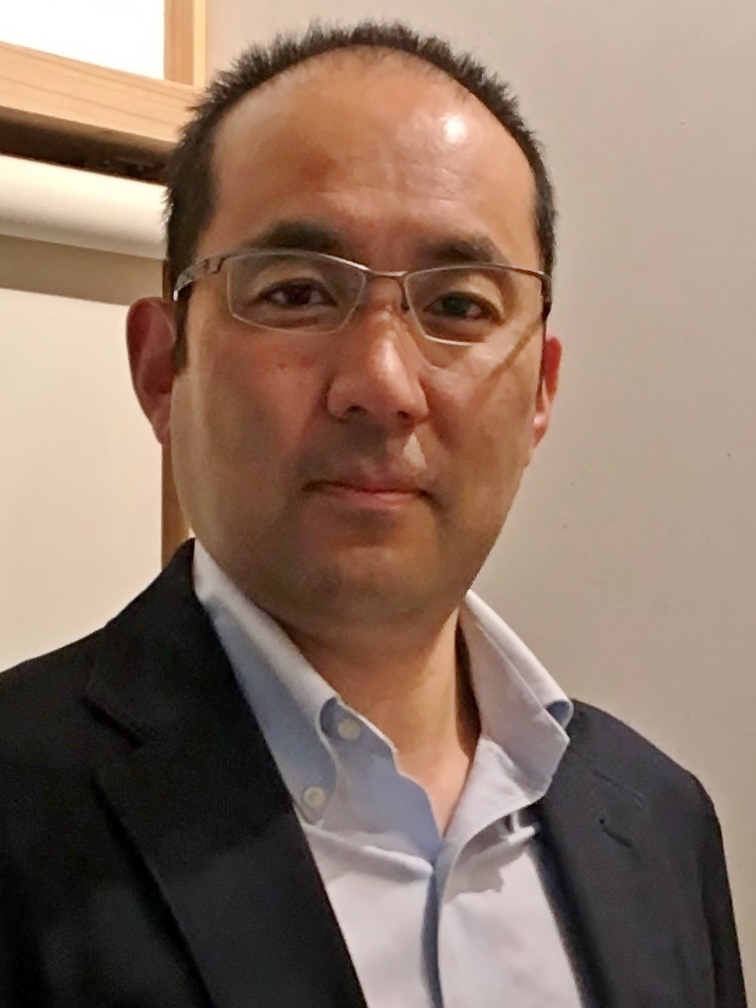}}]
{Takahiro~Wada} 
(M'99) received a B.S. degree in Mechanical Engineering, a M.S. degree in Information Science and Systems Engineering, and a Ph.D. degree in Robotics from Ritsumeikan University, Japan, in 1994, 1996, and 1999, respectively. 
He as a Research Associate worked at Ritsumeikan University (1999--2000).
He worked at Kagawa University as a Research Associate (2000--2003), an Assistant Professor (2003--2007) and Associate Professor (2007--2012).  
He has been a full professor at Ritsumeikan University (2012--2021) and at Nara Institute of Science and Technology (2021--present).
His research interests include robotics, human-machine systems, and motion sickness modeling.
He is a member of IEEE (RAS, ITSS, SMC), SAE, HFES, SICE, JSAE, RSJ, JSME.
\end{IEEEbiography}

\end{document}